\documentclass[12pt]{article}
\usepackage[margin=2.5cm]{geometry}  
\usepackage{epsfig, graphics}
\usepackage{longtable}
\usepackage{xcolor}
\usepackage{amsmath}
\usepackage[utf8]{inputenc} 
\usepackage{accents}
\usepackage{subcaption}
\usepackage{booktabs,caption}
\usepackage[flushleft]{threeparttable}
\usepackage{caption}
\usepackage{float}
\usepackage{bm}
\usepackage{amsmath}
\usepackage{natbib}
\usepackage{booktabs}
\usepackage{amsfonts}
\usepackage{amssymb}

\usepackage{ltxtable, filecontents}
\usepackage{tabularx}
\usepackage{setspace}
\usepackage{amsmath}
\newcommand{\nin}{\noindent}

\def\bkR{{\rm I\kern-.17em R}}

\def \id{1\hskip -3pt \mbox{l}}
\def \1n{1\hskip -3pt \mbox{N}}

\newfont{\bbf}{cmbx12 scaled 1435}

\begin{document}
\setlength{\baselineskip}{.26in}

\thispagestyle{empty}
\renewcommand{\thefootnote}{\fnsymbol{footnote}}
\vspace*{0cm}
\begin{center}

\setlength{\baselineskip}{.32in}
{\bf Intraday Functional PCA Forecasting of Cryptocurrency Returns}\\

\setlength{\baselineskip}{.26in}
\vspace{0.4in}
Joann Jasiak\footnote[1]{York University, Canada, {\it e-mail}:
{\tt jasiakj@yorku.ca}},  Cheng Zhong\footnote[2]{York University, Canada, {\it e-mail}:
{\tt cz1989@my.yorku.ca} \\
The authors gratefully acknowledge the financial support of the Natural Sciences and Engineering Council (NSERC) of Canada and thank C. Gourieroux for helpful comments.\\
This paper has been presented at 2025 CES North American Conference, University of Michigan's Ross School of Business and the Canadian Economic Association (CEA) meetings, Montreal 2025}

\today\\

\medskip

\vspace{0.3in}
\begin{minipage}[t]{12cm}
\small
\begin{center}
Abstract \\
\end{center}
We study the Functional PCA (FPCA) forecasting method in application to functions of intraday returns on Bitcoin. We show that improved interval forecasts of future return functions are obtained when the conditional heteroscedasticity of return functions is taken into account. 
The Karhunen-Loeve (KL) dynamic factor model is introduced to bridge the functional and discrete time dynamic models. It offers a convenient framework for functional time series analysis. For intraday forecasting, we introduce a new algorithm based on the FPCA applied by rolling, which can be used for any data observed continuously 24/7. The proposed FPCA forecasting methods are applied to return functions computed from data sampled hourly and at 15-minute intervals. Next, the functional forecasts evaluated at discrete points in time are compared with the forecasts based on other methods, including machine learning and a traditional ARMA model. The proposed FPCA-based methods perform well in terms of forecast accuracy and outperform competitors in terms of directional (sign) of return forecasts at fixed points in time.  \\


\end{minipage}

\end{center}
\renewcommand{\thefootnote}{\arabic{footnote}}

\newpage

\section{Introduction}

Bitcoin, the world's first cryptocurrency introduced in 2008, has become one of the most important digital assets worldwide. Its decentralized, peer-to-peer network, underpinned by blockchain technology, ensures transparency, security, and independence from traditional banking systems. Like other cryptocurrencies, Bitcoin is traded 24 hours a day, 7 days a week. 

Recently, Bitcoin prices have increased considerably and its returns remained highly volatile. The high return volatility and speculative nature of Bitcoin present both opportunities and challenges for investors. Bitcoin's price dynamics are influenced by a complex interplay of factors, including market sentiment, regulatory developments, technological advancements, macroeconomic trends, and geopolitical events. These factors introduce substantial uncertainty, making Bitcoin forecasting particularly challenging.

Several studies have employed various methods to forecast Bitcoin returns and volatility. \citet{katsiampa2017volatility} applied traditional time series models, such as GARCH, to estimate Bitcoin's volatility, providing a foundational approach to understanding its price dynamics. \citet{bouri2017return} explored the return-volatility relationship in the Bitcoin market, highlighting the asset's unique risk-return profile. 
\citet{gradojevic2023forecasting} compared the performance of Feedforward deep artificial neural network(FF-D-ANN), support vector machine(SVM) and random forest(RF) in predicting daily and hourly Bitcoin returns from two linear models: random walk (RW) and ARMAX(1,1). 

Since Bitcoin is traded continuously, its returns on a given day can be viewed as functions of time measured in minutes or seconds of UTC\footnote{Coordinated Universal Time} [see \citet{zhong2024} for Bitcoin returns, and e.g. \citet{AueAlexander2017FGAC}, \citet{KokoszkaPiotr2017Ifta} for applications to intraday returns on stocks, stock market indexes, including  S$\&$P500 and ETF's]. Then, the daily return functions can be modeled as a functional time series and predicted, using the Functional Autoregressive process or order 1 (FAR(1)). \citet{aue2015prediction} show that this approach is equivalent to predicting from the Karhunen-Loeve expansion of a functional time series and Vector Autoregressive of order 1 (VAR(1)) model of eigenscores obtained from a Functional Principal Component Analysis (FPCA) of the covariance matrix. The advantage of the latter approach is that it can easily be adapted to the return dynamics characterized by conditional heteroscedasticity. The extension of the FPCA-based model to conditionally heteroscedastic data is the main objective of this paper.

We introduce a new Functional Principal Component Analysis (FPCA)-based method for forecasting daily Bitcoin return functions out of sample. 
Our approach aims at improved interval forecasting of daily functions of returns and exploits the serial correlation of eigenscores and squared eigenscores. 
The novelty of our approach is that it  takes into consideration the conditional heteroscedasticity of eigenscores and produces forecast intervals that are more accurate and narrower than those obtained when the conditional heteroscedasticity is disregarded. Since the functional dynamic models for serially correlated and conditionally heteroscedastic functional time series are either complicated, or even not available in the literature, we introduce the Karhunen-Loeve (KL) dynamic factor model by imposing simplifying assumptions on the Karhunen-Loeve series expansion. This allows us to bridge the conditionally heteroscedastic dynamic discrete time dynamic models with their functional counterparts. Moreover, the KL dynamic factor model offers a convenient framework for the analysis of conditionally heteroscedastic functional time series.

Our second contribution is a rolling FPCA that allows for intraday forecasting of Bitcoin returns as subsets of a daily function. We observe that Bitcoin and many other financial assets are traded 24/7, and the delimitation of the day between hours 0:0 and 24:00 of UTC is purely conventional. This motivates us to consider the FPCA of partially overlapping daily return functions that start at subsequent time points to forecast a future segment of an incomplete function of current intraday returns.

The advantage of the functional approach compared to the conventional forecasting of discrete data is that the function can be evaluated at any time and provide forecasts at any point of time during the day rather than at a fixed sampling frequency.

In our empirical study, we examine a sample of daily functions of 15 minute and hourly returns. We model and forecast the return functions and evaluate the performance of our forecasting methods. Our method of forecasting daily return functions outperforms the FPCA-based forecasting methods available in the literature in that it produces more accurate sign forecasts and narrower forecast intervals. Our proposed algorithm for intraday return forecasting based on a rolling FPCA is shown to outperform the machine learning methods in terms of forecast accuracy and sign at fixed points of time.

In the literature, the Functional Principal Component Analysis (FPCA) has been used for forecasting first by 
\citet{aguilera1997approximated}, who proposed a principal component prediction model using functional principal component analysis for continuous time stochastic processes. The method is applicable to forecasting long segments of future functions so that the functional time interval can be divided into the associated "past", i.e. t=1,..., T and "future", i.e. t=T+1,...,T+K, say. The FPCA is then applied separately to the divided functional data relating to the "past" and "future" times, and the eigenscores and eigenfunctions are computed for the "past" and "future" independently. The forecast future values of the function are estimated from a linear combination of the estimated "future" eigenscores and "past" eigenfunctions plus the estimated future mean.
Our intraday FPCA forecasting is in part inspired by this method and allows for forecasting in real time, with the "future" potentially reduced to a neighborhood of one point in time.

\citet{aue2015prediction} proposed an approach to functional forecasting of complete future functions using the FPCA combined with a Vector Autoregressive (VAR(1)) model of eigenscores. This method was applied to environmental data that follow a stationary (non-trending) process of pollution concentrations, specifically focusing on forecasting half-hourly measurements of PM10 concentrations in ambient air in Graz, Austria, from October 1, 2010, to March 31, 2011. \citet{shang2020dynamic} used a similar approach, except that they used univariate autoregressive-moving average ARMA(p,q) time series models to forecast each eigenscore process independently in a study of Japanese age-specific mortality rates obtained from the Japanese Mortality Database.  \citet{shang2022dynamic} also applied independently the univariate ARMA(p,q) time series models to forecast the eigenscores in an empirical study of implied volatility of foreign exchange rates.

Our functional return forecasts improve upon the methods of \citet{aue2015prediction}, \citet{shang2020dynamic} and \citet{shang2022dynamic} by adjusting them to the dynamics of financial returns. In particular, we observe that the eigenscores of returns are characterized by serial correlation at short lags, which can be efficiently modeled by the Autoregressive AR(1) process rather than an ARMA(p,q). Most importantly, we evidence serial correlation in the squared eigenscores, i.e. we show that they are conditionally heteroscedastic and characterized by the ARCH effect. This characteristic is important, and by accounting for the conditional heteroscedasticity of eigenscores, we produce improved pointwise interval forecasts of return functions. 

In addition, to account for lagged cross-correlations between the eigenscores and their squares, we improve the approach of \citet{aue2015prediction} by replacing the VAR(1) forecast of eigenscores, by the VAR(1)-sBEKK(1,1) model that accounts for serial correlation and conditional heteroscedasticity of eigenscores in a multivariate setup. This approach provides an additional improvement in terms of interval forecast accuracy, as the pointwise VAR-sBEKK interval forecast outperforms our AR-GARCH based method of forecasting univariate time series of eigenscores separately.

The paper is organized as follows: Section 2 describes the FPCA and KL dynamic models, introduces the new forecast method of the forecast interval and compares it with the existing FPCA forecasting models. Section 3 describes the functional data on Bitcoin and presents the results of our forecasting method for daily return functions. Section 4 introduces a new method of forecasting of returns on Bitcoin from a rolling FPCA when the function of observations is incomplete, and compares it with other Bitcoin return forecast models. Section 5 concludes the paper. Additional results are provided in the Appendices A and B..

\section{Methodology}

This section reviews the Functional Principal Component Analysis (FPCA), introduces the new approach to functional times series forecasting of conditionally heteroscedastic returns, and compares it with the existing methods.

\subsection{The Functional Principal Component Analysis (FPCA)}

\medskip

In the literature [see, e.g \citet{ramsay2005principal}] functional data are usually modeled as "noise-corrupted observations on a set of trajectories" that are assumed to be realizations of a smooth random function of time $X(t)$, with unknown mean shape $\mu(t)$ and covariance $c(t,s) = cov(X(t),X(s))$ \citep{fpcatheoryliu2017functional}.

Let $\{X_i(t)\}_{i \in \mathbb{Z}}$ be a stationary functional time series, where $t \in I = [a,b]$ is a continuum, and $i=1,2,...$ denote the realization. Each realization can be written as a function $X_i(t, \omega)$ in both time $t \in I$ and $\omega \in \Omega$, an element of probability space. For a fixed $\omega$, and for each $i$, $X_i(t)$ is in the Hilbert space $H = l^2(I)$, equipped with the inner product $\langle x,y\rangle_{l^2} = \int_I x(t) y(t)dt$ of square-integrable real functions on a support $I \subset \mathbb{R}$  and the norm $||x||^2_{l^2} = \int_I x^2(t) dt$ \citep{loeve1978},
\citep{shang2022dynamic}. At a given $t$ and for each $i$, $X_i(t, \omega)$ is a stochastic variable defined
on a common probability space with finite second-order moment, i.e. any $X_i(t)$ is such that $E X_i(t)^2<\infty$ $\forall i,t$, where $E(X_i(t)^2) = \int_{\Omega} X_i^2(t, \omega) dP(\omega)$. The space $L^2(\Omega, \cal{A}, P)$ of variables with finite second-order moments is another Hilbert space equipped with the inner product $<X,Y>_{L^2} = \int_{\Omega} X(\omega) Y(\omega) dP(\omega)$.


\textbf{Karhunen-Lo\`eve Expansion}

The Karhunen-Lo\`eve expansion can be written for each stochastic function $X_i = \{ X_i(t),$ $t \in [a,b] \}$ that is zero-mean, or demeaned, i.e. replaced by $X_i(t) - \mu(t)$, and where $\mu(t) = E X_i(t)$ is independent of index $i$ by the stationarity assumption.


\nin The covariance function $c(t,s)$ is:
\begin{equation}
c(t,s) = E{[X(t) - \mu(t)][X(s) - \mu(s)]}, \; \forall t,s \in I, \label{cov}
\end{equation}

\nin where $\mu(t)$ is the mean function. The covariance function $c(t,s)$ allows the covariance covariance operator of $X$ denoted by $K$ to be defined as:
\begin{equation}
 K(\xi)(s) = \int_I c(t,s) \xi(t) dt. 
\end{equation}

\nin This operator is a function from $l^2(I)$ onto $l^2(I)$ that associates the element $K(\xi) \in l^2(I)$ with any element $\xi$ of $l^2(I)$. This operator is continuous, symmetric, positive, i.e. $<\xi, K(\xi)>_{l^2(I)} \geq 0$, $\forall \xi \in l^2(I)$.

It follows from Mercer's and Karhunen-Lo\`eve  Theorems that there is an orthonormal sequence $(\xi_j)$ of  functions in $l^2(I)$ and a non-increasing sequence of positive numbers $(\lambda_j)$ such that the operator $K$ admits a spectral decomposition: 

\begin{equation}
    c(t,s) = \sum_{j=1}^\infty \lambda_j \xi_j(t)\xi_j(s), \qquad t,s \in I
\end{equation}

\nin where $\lambda_j, j=1,2,...$ are the eigenvalues in strictly decreasing order and $\xi_j(s), j=1,2,.. $ are the normalized eigenfunctions, so that $K(\xi_j) = \lambda_j \xi_j$  with  $< \xi_j, \xi_k>_{l^2} \equiv \delta_{j,k} = 1$, if $j=k$, and $0$, otherwise. Moreover, $\xi_j$ are continuous functions if the associated $\lambda_j$ are strictly positive and the equality corresponds to both a convergence in $L^2$ for the Karhunen-Loève Theorem and a uniform convergence in Mercer's Theorem.

Then, each stochastic function $X_i(t, \omega)$, $i=1,2,...$ can be expressed as a linear combination of these basis functions. Through Karhunen-Loève expansion, a stochastic function $X(t, \omega)$  can be expressed as a linear combination of orthogonal basis functions $\xi_j$ determined by the covariance function of the process. 

$$ X(t,\omega ) = \mu(t) + \sum_{j=1}^\infty \beta_{j} (\omega) \xi_j(t),$$

\nin or, by omitting $\omega$ for ease of exposition, as:

\begin{equation}
X(t)  = \mu(t) + \sum_{j=1}^\infty \beta_{j} \xi_j(t),
\end{equation}

\nin where $\beta_{j}$ are eigenscores obtained from a projection in $l^2(I)$ of $[X(t)-\mu(t)]$ in the direction of the $j$th eigenfunction $\xi_j$, i.e. $\beta_{j} = <[X(t) - \mu(t)], \xi_j(t)>_{l^2(I)}$.  The eigenscores $\beta_j, \; j=1,2,.. J $ are pairwise uncorrelated random variables with zero mean and variance $\lambda_j$ \citep{benko2009}. Each of them can be interpreted as the contribution of $\xi_j(t)$ to  $X(t)-\mu(t)$.  The functions $\xi_j$ are continuous real-valued deterministic functions on $I=[a,b]$ that are pairwise orthogonal in $l^2(I)$, i.e. in the time domain. If $X(t)$ is Gaussian, then the random variables $\beta_j$ are also Gaussian, uncorrelated, and then stochastically independent. 

Alternatively, we can write:

$$X(t, \omega)  = \mu(t) + \sum_{j=1}^\infty \sqrt{\lambda_j} \beta_{j}^*(\omega) \xi_j(t),$$

\nin to see that $\beta^*_j(\omega)$ are orthonormal in $L^2(\Omega, \cal{A}, P)$, and $\xi_j$ are orthonormal in $l^2(I)$.  It is therefore sometimes said that the Karhunen-Loève expansion is bi-orthogonal (bi-orthonormal).

\subsection{The KL Dynamic Factor Model}

We consider the stationary functional time series indexed by $i=1,...,N$ and the Karhunen-Loève (KL) expansion of $X_i(t)$ for each $i$:

\begin{equation} 
X_i(t)  = \mu(t) + \sum_{j=1}^\infty \sqrt{\lambda_j} \beta_{i,j}^* \xi_j(t),
\end{equation}

\nin where the stochastic eigenscores $\beta_{i,j}^*$  are zero-mean with unit variance, uncorrelated for different $j$, stationary in $i$, but can be correlated for different $i$.

\subsubsection{The Model}

The above expansion is an equivalent representation of a stationary functional process (stationary in $i$) and imposes no other restriction, except for some weak regularity conditions. It can be written as:

\begin{equation} 
X_i(t)  = \mu(t) + \sum_{j=1}^J \sqrt{\lambda_j} \beta_{i,j}^* \xi_j(t) + e_i(t),
\end{equation}

\nin where $e_i(t) = \sum_{j=J+1}^{\infty} \sqrt{\lambda_j} \beta_{i,j}^* \xi_j(t)$.
The truncation error function, denoted by $e_i(t)$, is stationary, with mean zero and finite variance. The tuning parameter $J$ represents the number of eigenfunctions that are preserved in the approximation, resulting in a dimension reduction \citep{shang2022dynamic}.

The representation (4) or (6) can be used to define a structural model for functional time series by specifying the dynamics of the eigenscores, either
$\beta_{i,j}^*$, or $\beta_{i,j} = \sqrt{\lambda}_j \beta_{i,j}^*$. More precisely, we make the following assumption:

\medskip

\nin \textbf{Assumption 1:}

i) There exists a true value $J_0$ of $J$.

ii) The eigenscore series $\beta_{i|J_0} = (\beta_{i,1},...,\beta_{i,J_0})$ and
$\beta_{i|\bar{J}_0} = (\beta_{i,J_0+1},\beta_{i,J_0+2},...)$ are independent.

iii) The conditional distribution of $\beta_{i|J_0} = (\beta_{i,1},...,\beta_{i,J_0})$
given $\beta_{\underline{i-1|J_0}} = (\beta_{i-1|J_0},\beta_{i-2|J_0},...$  has a parametric specification with parameter $\theta$.

iv) The remaining $\beta_{i|\bar{J}_0} = (\beta_{i,j| j > J_0})$ are independent white noises.

\medskip

\nin \textbf{Definition 1:} The KL dynamic factor model satisfies the decomposition (6) and Assumption 1. 

\medskip

\nin It is easily checked that the KL expansion appears as a dynamic factor model with $J_0$ dynamic factors $\beta_{i|J_0}, i=1,...,J_0$ and an independent and identically distributed (i.i.d.) functional noise $e_i(t)$. The noise is partly degenerate since by construction, we have:

$$< e_i, \xi_j>_{l^2(I)} = 0, \;\forall j=1,...,J_0,$$

\nin and the support of its distribution is in the vector space orthogonal to the space of $\xi_j, j=1,...,J_0$ in $l^2(I)$.

\medskip
\nin {\it Example 1: KL Linear Dynamic Factor Model}

For ease of exposition we consider a Gaussian framework with linear dynamics of order 1. Then, the specification is:

\begin{equation}
X_i(t) = \mu(t) + \sum_{j=1}^{J_0} \beta_{i,j} \xi_j(t) + e_i(t),
\end{equation}

\nin where

\begin{equation}
\beta_{i|J_0} = \Phi \beta_{i-1| J_0} + \epsilon_{i},
\end{equation}

\nin where $\beta_{i|J_0}$ is a vector of length $J_0$ containing $\beta_{i,j}, j=1,...,J_0$, the autoregressive matrix $\Phi$ of dimension $J_0 \times J_0$ has eigenvalues inside the unit circle, $(\epsilon_i), i=1,2,..$ is an independent and identically distributed (i.i.d.) white noise of dimension $J_0$, $\epsilon_i \sim N(0, \Sigma), \forall i$, assumed to be independent of the functional noise $\{e_i(t), t \in [a,b]\}, i=1,...,N$. The functional noise has a functional Gaussian distribution of infinite dimension denoted by $N(0, \sigma^2 \Omega_e)$, where 
$\Omega_e = Id - \sum_{j=1}^{J_0} \xi_j \xi_j'$, and $Id$ denotes the Identity operator on $l_2(I)$ and 
$\xi_j' \Omega_e \xi_j =0, j=1,...,J_0$.
The model (7)-(8) depends on the parameters of different types. Some of them are scalar or multidimensional parameters, such as $\Phi, \Sigma, \sigma^2$, $\lambda_j, j=1,...,J_0$,  while $\mu, \xi_j, j=1,...,J_0$, are functional parameters. 

\medskip
The above model can be viewed in another way. It is equivalent to consider the functional series $X_i(t), t \in I$, or the countable set of series $\beta_{i,N}
= (\beta_{i,1},...,\beta_{i,j},...)$
that are related by a "one-to-one" linear relationship. Therefore, it is equivalent to write a Gaussian autoregressive model for the $X$'s or for the $\beta$'s. In the framework of this example, we have introduced a constrained Gaussian VAR(1) specification for the $\beta$'s. Therefore, a reduced-rank, constrained functional Gaussian FAR(1) is the associated representation of the functional process.

Let us now justify this representation, as a convenient specification of the dynamics of $\beta$'s for the Functional Vector Autoregressive
FAR(1) process:

\begin{equation}
X_i - \mu = \Psi(X_{i-1} -\mu) + u_i,
\end{equation}

\nin where the time $t$ is omitted for ease of exposition. Under the approach of \citet{2000far1} the errors $(u_i, i \in \mathbb{Z})$ are centered (i.i.d.) innovations in $l_2(I)$ satisfying $E ||u_i||^4 = E[\int u_i^2(t)dt]^2 < \infty$
and $\Psi$ is a bounded linear operator satisfying the condition
$\int \int \psi^2(s,t) ds dt <1$ that ensures that the above equation has a strictly stationary and causal solution such that $u_i$ is independent of $X_{i-1}, X_{i-2},..$.

Suppose, $\mu=0$ and $J_0$ is given by Assumption 1 i). Then, it follows from Aue at al (2015), Appendix A.1 that for $J=J_0$ the process can be written as
\begin{eqnarray}
<X_i, \hat{\xi}_j> & = & < \Psi(X_{i-1}), \xi_j> + <u_i, \xi_j> \nonumber \\
& = & \sum_{j'=1}^{\infty} < X_{i-1}, \xi_{j'} > < \Psi(\xi_{j'}), \xi_j> + < u_{i},\xi_j>   \nonumber \\
& = & \sum_{j'=1}^{J_0} < X_{i-1}, \xi_{j'} > < \Psi(\xi_{j'}), \xi_j  > + \epsilon_{i,j} \\ \nonumber
\end{eqnarray}
\nin where $\epsilon_{i,j} = d_{i,j} + < u_{i},\xi_j>$ with the remainder terms
$d_{i,j} = \sum_{j'=J_0+1}^{\infty} < Y_{i-1}, \xi_{j'} >
< \Psi( \hat{\xi}_{j'}), \hat{\xi}_j  >$ being independent white noises by Assumption 1 iv).

Next, the empirical counterpart of the above inner product $\beta_i = <X_i, \hat{\xi}_j>$ is written as follows.  Let \(\mathbf{\beta}_{i} = (\beta_{i,1}, \beta_{i,2}, \dots, \beta_{i,J_0})'\) be a $(J_0 \times1) $ vector of eigenscores on day $i$, and write

\begin{equation}
    \boldsymbol{\beta}_i = \boldsymbol{\Pi}_1\boldsymbol{\beta}_{i-1} + \boldsymbol{\epsilon}_i
    \label{var}, \; i=2,...N
\end{equation}

\nin where $\mathbf{\Pi_1}$ is the coefficient matrix of dimension $(J_0 \times J_0)$ with eigenvalues of modulus strictly less than 1,  and $\boldsymbol{\epsilon}_i$ is a multivariate $(J_0 \times 1)$ white noise process with mean zero and variance $\Sigma $.
 Note that the marginal variance of each eigenscore vector being equal to an Identity matrix, implies an additional constraint $\Sigma = Id - \Pi \Pi'$ on the parameters of the model.

\medskip
\nin {\it Example 2: KL Conditionally Heteroscedastic Dynamic Factor Model}

Let us consider the model of Example 1, with equation (8) replaced by

\begin{equation}
\beta_{i|J_0} = \Phi \beta_{i-1, J_0} + \epsilon_i, \; \mbox{where} \; \epsilon_i = H_i^{1/2} \eta_{i},
\end{equation}

\nin and where $H_i$ is the conditional covariance matrix of $\beta_{i|J_0}$ of dimension $J_0 \times J_0$ given the information set ${\cal F}_{i-1}$ generated by the realizations up to and including $i-1$, and $\eta_i$ is an i.i.d. vector with mean 0 and such that $E\eta \eta' = Id$.


Then, a decomposition similar to (10) can be considered under a relaxed assumption on the errors
being  $L^4-m$-approximable \citep{KokoszkaPiotr2017Ifta}
leading to $\epsilon_{i,j}$, elements of vector $\epsilon_i = (<u_i, \tilde{\xi}_{1}>,...,<u_i, \tilde{\xi}_{J_0}>)' + \mbox{white noise}$, which  
is a vector of linear transformations of the conditionally heteroscedastic error plus noise. Therefore it is heteroscedastic too. 

It is easy to see that in this context, a complicated multivariate functional ARCH model would be needed to represent the dynamics of conditional variances and the lagged cross-effects of the squared values of error functions. To our knowledge, such a multivariate functional model is not available. In contrast, the multivariate dynamics of conditional variance can be easily specified for the eigenscores in the context of the KL factor model, as follows.

The empirical counterpart of $\beta_i = <X_i, \hat{\xi}_j>$  can be specified as the VAR(1) process in eq. (11) 
with weak white noise errors and, for example, the BEKK model of conditional volatility  \citet{EngleRobertF.1995MSGA}. The BEKK model gets computationally costly when the number of time series data increases. Therefore, we can use its  variant, called the scalar BEKK (sBEKK(1,1)) \citep{DingZhuanxin2001LSCC}:
\begin{equation}
    \mathbf{H}_i = \mathbf{CC}' + a \boldsymbol{\epsilon}_{i-1}\boldsymbol{\epsilon}'_{i-1} + g\mathbf{H}_{i-1}
    \label{equ: bekk}, \; i=1,...N
\end{equation}
\nin where, $\mathbf{H}_i$ is the $(J_0 \times J_0)$ multivariate conditional variance-covariance matrix of $\beta_{i}$, $a,g \in R$ with $a\geq 0,g \geq 0$ are scalar parameters, and $\boldsymbol{\epsilon}_i$ is the error vector of the VAR(1) representation of eigenscores in equation (\ref{var}). Matrix $\mathbf{C}$ is a lower triangular matrix of parameters of dimension $J_0 \times J_0$. We assume that the parameters of the sBEKK(1,1) model satisfy the standard stationarity and non-negativity conditions. Since the marginal variance of each eigenscore vector is equal to an Identity matrix of dimension $J_0$, the parameters of this model are constraint too, as illustrated in Appendix B.

\medskip
\nin {\it Example 3: KL Conditionally Heteroscedastic Univariate Dynamic Factor Model}

The KL factor model can be simplified further, as in practice one may prefer to model the contemporaneously uncorrelated eigenscore processes using either univariate or multivariate time series models \citep{shang2020dynamic, shang2022dynamic}. In addition, for conditionally heteroscadastic processes we can specify the GARCH dynamics for the associated errors.

Let us suppose that the eigenscores $\beta_{i,1},..,\beta_{i,J_0}$ are independent and follow univariate AR($p$)-GARCH($K,L$) models. Then, the dynamic of eigenscore $\beta_{i,j}$ can be written as:

\begin{align}
\beta_{i,j} &= \sum_{l = 1}^p a_{l,j} \beta_{i-l,j} + \epsilon_{i,j},  \\
    \nu_{i,j} &= \varsigma_{0,j} + \sum_{k = 1}^K \varsigma_{k,j} \nu_{i-k,j} + \sum_{l=1}^L \zeta_{l,j} \epsilon_{i-l,j}^2, \nonumber
\end{align}

where
$$\epsilon_{i,j} = \sqrt{v_{i,j}} z_{i,j}, \;\; \forall i=1,..,N, \\ $$

\nin and where $a_{l,j} $ is the autoregressive coefficient on lag $l$ of eigenscore $j$,  $\epsilon_{i,j}$ is the error term of eigenscore $j$, $\varsigma_0, \varsigma_{l,j}$ and $\zeta_{k,j}$ are the coefficients of the conditional variance,  and $z_{i,j}$'s are i.i.d. N(0,1) variables \citep{gourieroux2012arch}. We assume that the roots of the autoregressive polynomials $1-\sum_{l = 1}^p a_{l,j} x^l=0$ are outside the unit circle $\forall j=1,..,J_0$, $\varsigma_0>0$, and the remaining coefficients of the GARCH process satisfy the standard stationarity and non-negativity conditions. Note that the marginal variance of each eigenscore being equal to 1, implies additional constraints on the parameters of the model, which are illustrated in Appendix B.

\subsubsection{The KL Dynamic Factor-based Functional Forecast}

Let us assume a KL dynamic factor model with mean zero satisfying Definition 1. The theoretical out-of-sample forecast and the conditional variance are derived below in the general case that includes Examples 1,2, and 3.
$$
  E(X_{N+1}|  \underline{X_{N}})  = E([\sum_{j=1}^{J_0} \beta_{N+1,j} \xi_j]| \underline{X_{N}}) 
  = \sum_{j=1}^{J_0} [ E( \beta_{N+1,j}|\underline{X_{N}})] \xi_j,
$$

 \nin where $\underline{X_N}$ denotes the past curves and  $E( \beta_{N+1,j} \underline{ X_{N}})$ is the forecast of the random eigenscore.
By definition, the prediction of $\beta_{N+1,j}$ depends on the past curves through the $J_0$ first eigenscores only. Therefore, we have:

$$E(X_{N+1}|  \underline{X_{N}})  =  \sum_{j=1}^{J_0} [ E( \beta_{N+1,j}|\underline{\beta_{N|J_0}})] \xi_j.$$

\nin It is either based on $E( \beta_{N+1,j}| \underline{\beta_{N,j}})$ when each eigenscore $\beta_{N+1,j}, \; j=1,..,J_0$ is predicted from its own past based on a univariate time series model (Example 2), or $E( \beta_{N+1,j} | \underline{\beta_{N}})$ where $\beta_{N}=(\beta_{N,1},...,\beta_{N,J_0})$ when the eigenscore $j$ is predicted from its own past and the past values of the remaining $J_0-1$ eigenscores (Examples 1 and 3).

The conditional variance at horizon 1 is:
\begin{equation}V(X_{N+1}|\underline{X_N}) = \sum_{j=1}^{J_0} \sum_{l=1}^{J_0} \xi_j \xi_l' Cov(
\beta_{N+1,j}, \beta_{N+1,l}|\underline{\beta_{N|J_0}}) + \omega,
\end{equation}

\nin where $\omega= \sigma^2 \sum_{j=J_0+1}^{\infty} \xi_j \xi_j'$, $\sum_{j=J_0+1}^{\infty} \xi_j \xi_j' = Id - \sum_{j=1}^{J_0} \xi_j \xi_j'$ where $Id$ denotes the identity operator on $l_2(I)$ and 
$\xi'$ denotes the adjoint operator of $\xi$.
Note that $\sum_{j=J_0+1}^{\infty} \xi_j \xi_j'$ and $\sum_{j=1}^{J_0} \xi_j \xi_j'$ are orthogonal projectors (symmetric and idempotent) in the KL dynamic factor model.

\subsection{New Approach to FPCA Forecasting of Financial Returns}

Suppose the returns on Bitcoin (with a fixed holding period) is a stationary time series of functions $X_i(t), t \in I$, each defined over one day and observed on consecutive days $i=1,...,N$. Then, the intraday Bitcoin returns can be treated as a continuous time function.

\subsubsection{Estimation}

\nin Wee approximate the expansions of stochastic functions in (4) by

\begin{equation}
X_i(t) = \mu(t) + \sum_{j=1}^J \beta_{i,j} \xi_j(t) + e_i(t), i=1,...,N.
\end{equation}

\nin The mean $\mu(t)$ can be consistently estimated from a sample of $N$ functional observations as $\hat{\mu}_N(t) = \frac{1}{N} \sum_{i=1}^N X_i(t)$. The covariance operator $K$ can be consistently estimated from $\hat{c}_N(x) = \frac{1}{N} <X_i -\hat{\mu}_N,x> (X_i - \hat{\mu}_N)$  \citep{ramsay2005principal}. \citet{aue2015prediction} show that under general weak dependence assumptions these estimators are $\sqrt{N}$ consistent. \citet{KokoszkaPiotr2017Ifta} show the consistency and normality of the mean and covariance estimators for functional time series that are autocorrelated and conditionally heteroscedastic. 

Under the static FPCA, the functional principal components are extracted from $c(t,s)$, the estimated $\hat{K}$ yielding the estimated $\hat{\lambda}_j, \hat{\xi}_j, j=1,...,J$, where
$[\hat{\xi}_1(t),  \hat{\xi}_2(t),...]$ are the orthogonal sample eigenfunctions obtained from  $\hat{c}(s,t) = \sum_{j=1}^J \hat{\lambda}_j \hat{\xi}_j(t) \hat{\xi}_j(s)$ where $\hat{\lambda}_1 > \hat{\lambda}_2  > \cdots \geq 0$ are the sample eigenvalues of $\hat{c}(s,t)$ under the static approach. 
The static FPCA approach has been extended to dynamic FPCA by \citet{aue2015prediction} for serially correlated functions. Under the dynamic FPCA, the functional principal components are extracted from the long-run covariance matrix defined as $C(t,s) = \sum_{l=-\infty}^{\infty} cov (X_0(s), X_l(t))$, i.e. the marginal covariance of the stationary time series of functions, to ensure the consistency of the estimates. \citet{shang2022dynamic} observe that the estimation of dynamic functional principal components using long-run covariance benefits the forecast if there are temporal dependencies in the functional data. However, \citet{shang2020dynamic} claim that the functional time series method with dynamic functional principal component decomposition does not always outperform that with static
functional principal component decomposition.

After the estimation,
the in-sample fitted functions $\hat{X}_i(t)$, $i=1,...,N$ are obtained from the sample of functional observations $X_1(t)$ $,....,X_N(t)$ as:
\begin{align}
\hat{X}_i(t)& = \hat{\mu}(t) + \sum_{j=1}^J \hat{\beta}_{ij} \hat{\xi}_j(t),\; i=1,...,N
\end{align}

\nin where $\hat{\mu}(t) = \frac{1}{N} \sum_{i=1}^N X_i(t)$ is the estimated mean function, $\hat{\beta}_{ij}$ is the $j^{th}$ estimated principal component score for the $i^{th}$ observation where $\hat{\beta}_i = < [X(t) - \hat{\mu}(t)] \hat{\xi}_i(t)>$.

In practice, the optimal choice of truncation $J$ is crucial for the estimation accuracy. The choice criteria are discussed in \citet{shang2022dynamic}, p.1028 and \citet{shang2020dynamic}, p.31. A commonly used approach chooses $J$ set at the minimum, allowing to reach a given fraction $\delta$ of the cumulative covariance explained by the first $J$ leading components. For example $J = argmin_{J:J\geq 1} \{\sum_{j=1}^J \hat{\lambda}_j/\sum_{j=1}^N \hat{\lambda}_j \id_{\hat{\lambda}_j>0} \geq \delta \}$ , with $\delta = 0.85 $.

This variance $\sigma^2$ of $e(t)$ can be estimated from the average of estimated daily variances of $\hat{e}_i(t) = X_i(t)-\hat{X}_i(t)$ over $N$ days, where $\hat{X}_i(t)$ is the fitted value defined in eq. (17). Next, to obtain $\hat{\omega}$, we multiply $\hat{\sigma}^2$ by $diag(Id - \sum_{j=1}^{J} \hat{\xi}_j \hat{\xi}_j')$, based on the entire sample of $N$ functions to approximate it according to eq. 15. We expect the result of $Id - \sum_{j=1}^{J} \hat{\xi}_j \hat{\xi}_j'$ to be numerically very close to an identity matrix. 

In practice the R-package \textit{fda} can be used to perform the FPCA, as it is done in this paper.
The estimation of the dynamic parameters of eigenscore models depends on the dynamic model chosen for forecasting of eigenscores, and it is discussed in the next subsection. 



\subsubsection{Forecast}
In general, the forecast at horizon $h$ is computed for as follows:

\begin{align}
\hat{X}_{N+h|N}& =  \hat{\mu} + \sum_{j=1}^J \hat{\beta}_{N+h|N,j} \hat{\xi}_j, 
\label{forecast function}
\end{align}

\nin where $\hat{\xi}= [\hat{\xi}_1,...,\hat{\xi}_J]$ is the set of estimated functional principal components estimated from $N$ functional observations and $\hat{\beta}_{N+h|N,j} $ is the forecast of the $j^{th}$ eigenscore. The forecast variance is
\begin{equation}V(\hat{X}_{N+h|N}) = \sum_{j=1}^{J} \sum_{l=1}^{J} \hat{\xi}_j \hat{\xi}_l' \widehat{Cov}(
\hat{\beta}_{N+h|N,j}, \hat{\beta}_{N+h|h,l}) + \hat{\omega}
\end{equation}

\nin 

\nin The forecast of the $j^{th}$ eigenscore is obtained from a time series forecasting method and motivated by \citet{aue2015prediction} and the related papers by \citet{shang2020dynamic} and \citet{shang2022dynamic} who documented the serial correlation in eigenscore processes $(\beta_{i,j}), i=1,2,...$.

\subsubsection{Forecast Based on Univariate Time Series Model of Eigenscores}

In this case, each future eigenscore $\hat{\beta}_{(N+1)j}, \dots, \hat{\beta}_{(N+h)j}$  for $j = 1, \dots, J$ is forecast by using the observations on the past values of each $j^{th}$ eigenscore $\hat{\beta}_{1j}, \dots, \hat{\beta}_{Nj}$ independently. We estimate the AR($p$)-GARCH($K,L$) model (14) in one step by the maximum likelihood\footnote{The R package  \textit{rugarch} is used to estimate the AR-GARCH in this paper.} to forecast each individual eigenscore $\beta_{(N+1)j}, \; j = 1,\dots, J$  and predict each eigenscore along with its future volatility $\nu_{(N+1),j}$.

The out-of sample forecast of eigenscore $\beta_{N+1,j}$ on day $i=N+1$ and the associated forecast of conditional variance lead to the $(1-\alpha)\%$ pointwise forecast interval calculated as:

\begin{equation}
    \hat{X}_{N+1}(t)  \pm z_{\alpha/2}\sqrt{\sum_{j = 1}^J \hat{\nu}_{N+1,j}\hat{\xi}^2_j(t) + \hat{\omega}(t)}
    \label{fun: garchci}
\end{equation}
\nin where $\hat{\nu}_{N+1,j}$ is the forecast variance of $jth$ eigenscore estimated by the AR-GARCH for the future function on day $N+1$, $\hat{\xi}^2_j(t)$ is the estimated eigenfunction and $ z_{\alpha/2}$ denotes the quantile of the standard Normal distribution.

\subsubsection{Forecast Based on Multivariate Time Series Model of Eigenscores}

For the joint forecast of eigenscores, we apply the sBEKK model  to forecast the future conditional volatility of eigenscores (${\beta}_{(N+1)}, \; j = 1, \dots, J_0$) on day $i=N+1$ in two steps. First, we estimate the VAR model (11) and predict the vector of future scores $\hat{\beta}_{(N+1)}$. Next, we apply the sBEKK(1,1) model (13) to the residuals  $\mathbf{\hat{e}}_i, \; i=1,...N$ of the VAR(1) model estimated in the first step  and forecast the conditional covariance matrix of eigenscores $H_{N+1}$ one day ahead\footnote{The R package \textit{BEKKs} is used to estimate the sBEKK(1,1) model in this paper.}.

The pointwise forecast intervals at level $(1-\alpha)\%$ for the returns one-day-ahead are  computed from the diagonal elements of matrix $\hat{\boldsymbol{\xi}}'(t)H_{N+1}{\hat{\boldsymbol{\xi}}}(t)$  as follows

\begin{equation}
     {\hat{\bf{X}}}_{N+1}(t) \pm z_{\alpha/2}\sqrt{{diag (\hat{\boldsymbol{\xi}}'(t)H_{N+1}{\hat{\boldsymbol{\xi}}(t))}} +\hat{\omega}(t)}
     \label{equ: bekkic}
\end{equation}

\nin where \(\mathbf{\hat{X}}_{N+1}\) is the forecast one-day-ahead function of returns forecast from the eigenscores $\hat{\beta}_{(N+1),j}, \;j = 1,\dots, J$, estimated by VAR(1)-sBEKK(1,1) model, and \(\hat{\boldsymbol{\xi}}(t)\) contains the estimated eigenfunctions $(\hat{\xi}_1(t),\dots, \hat{\xi}_{J_0}(t))'$

\subsubsection{Comparison with the Literature}

The main difference between our approach compared to \citet{aue2015prediction}, \citet{shang2020dynamic} and \citet{shang2022dynamic} is that the variance of the forecasted $j$th score is past dependent and predicted, instead of being considered constant and estimated ex-post as in step 4 of the algorithm of \citet{aue2015prediction}. In addition, the parameter estimates of the dynamic model of eigenscores account for conditional heteroscedasticity. Hence, they are more efficient, yielding more accurate forecasts of eigenscores.

In practice, the discrete time series of financial returns are characterized by conditional heteroscedasticity, in addition to potential serial correlation.
Then, such a time series is modeled as an autoregressive AR process with GARCH errors, i.e., the AR-GARCH process. For this reason,
we expect that the FPCA of return functions on Bitcoin generates eigenscores that are serially correlated and conditionally heteroscedastic.

Regarding the method of time series forecasting of eigenscores, \citet{aue2015prediction} use the Vector Autoregressive (VAR) model. It is motivated by the fact that although the scores are contemporaneously uncorrelated, there may exist cross-correlations at higher lags that are accounted for in a multivariate model.
Theorem 3.1 of  \citet{aue2015prediction} shows that the one-step ahead forecasts of functional time series from the KL expansion with VAR-based forecasts of eigenscores 
are asymptotically equivalent to the predictions based on the Functional Autoregressive process of order 1 (FAR(1)) fitted to the functional time series.

\citet{shang2020dynamic} and \citet{shang2022dynamic} use instead  independent univariate time series forecasting models for each $\beta_{N+h,j}, j=1,..., J$. Then, each future eigenscore $\hat{\beta}_{(N+1)j}, \dots, \hat{\beta}_{(N+h)j}$  for $j = 1, \dots, J$ is forecast by using the observations on the past values of each $j^{th}$ eigenscore $\hat{\beta}_{1j}, \dots, \hat{\beta}_{Nj}$ independently from the autoregressive moving average(ARMA)(p,q) model.
They report that the FPCA provides better out-of-sample forecasting performance than other functional models.

However, the ARMA model does not account for the potential presence of conditional heteroskedasticity, which can impact the accuracy of parameter estimators, the forecasts and the forecast intervals. The commonly used dynamic models for conditionally heteroskedastic discrete time series data are the Autoregressive and the Generalized Autoregressive Conditional Heteroskedasticity (ARCH and GARCH) models \citep{LIU2011724}. 
Modeling multivariate volatilities has played an important role in economics and finance studies. There are multivariate GARCH models (MGARCH) that exist in the literature, including the so-called vec-model of \citet{BollerslevTim1988ACAP}. However, this model does not guarantee a positive definite conditional covariance matrix. Other types of MGARCH models, such as the constant conditional correlation (CCC) model of \citet{BollerslevTim1990MtCi} and the dynamic conditional correlation (DCC) model of \citet{EngleRobert2002DCCA}, have been criticized in the literature too.

In the existing litrature, the forecast interval for FPCA models is obtained from the pointwise forecast bands algorithm. It was introduced by \citet{aue2015prediction}, and applied by \citet{shang2020dynamic} and \citet{shang2022dynamic}. It follows the steps given below, where for ease of exposition the horizon $h=1$ is considered.

\medskip
\nin \textbf{Step 1}: Estimate $J$ sample eigenscore vectors $(\hat{\beta_1}, \dots, \hat{\beta}_J)$ and sample eigenfunctions 

\nin $[\hat{\xi}_1(t), \dots, \hat{\xi}_J(t)]$ using all $N$ observations.  

\medskip
\nin \textbf{Step 2}: For $k \in [J+1, \dots, N-1]$, calculate the in-sample errors of the forecast function $$\hat{e}_{k+1}(t) =X_{k+1}(t) -\hat{X}_{k+1}(t), $$
where

\begin{equation*}
    \hat{X}_{k+1} = \sum_{j=1}^J\hat{\beta}_{(k+1)j} \hat{\xi}_j(t)
\end{equation*}
\nin and $\hat{\beta}_{(k+1)j}$ are forecasted by the chosen univariate or multivariate time series model. 

\nin \textbf{Step 3}: For each time point $t$, define the standard deviation of the in-sample errors
$\hat{e}_{k+1}(t)$ as: 

$$\gamma(t) = \sqrt{\frac{\sum\hat{e}_{k+1}(t)^2}{(N-1)-(J+1)}}$$. 

\nin \textbf{Step 4}: Seek turning parameters $\underline{\kappa}_\alpha, \bar{\kappa}_\alpha $ such that $\alpha \times 100\%$ of the errors satisfy

$$
-\underline{\kappa}_\alpha \gamma(t) \leq\hat{\epsilon}_k(t)\leq \bar{\kappa}_\alpha \gamma(t)
$$

\nin A practical approach to determine the pointwise forecast interval for a reasonable sample of size N, is:

\begin{align*}
        \frac{1}{N-J-1}\sum_{k = 1}^{N-J-1} &I(-\underline{\kappa}_\alpha\gamma(t) \leq \hat{\epsilon}_k(t)\leq\bar{\kappa}_\alpha\gamma(t)) \\
 &\approx P(-\underline{\kappa}_\alpha\gamma(t) \leq X_{k+1}(t) - \hat{X}_{k+1}(t)\leq\bar{\kappa}_\alpha\gamma(t))
\end{align*}

\nin for all points of time $t = 1, \dots, T$.  Compared with this approach, we use instead the  predicted volatility of eigenscores for improved forecast intervals.

\section{Data Analysis}

We examine return functions over one UTC day between hours 0:00 and 24:00, and computed from 15-minute and hourly data on prices from Bitsatmp, one of the largest cryptocurrency exchanges, covering the period from January 1, 2022, to December 30, 2023. The 15-minute data comprises 729 days with 96 observations per day, totaling 69984 observations of prices observed every 15 minutes and
17472 hourly prices. 

Below, we examine separately the return functions computed from data sampled at 15 minutes and hourly. 

\subsection{Intraday 15 Minute Returns}

\begin{figure}[H]
    \centering
        \caption{Functions of BTC 15-Minute Returns}
    \includegraphics[width=0.6\linewidth]{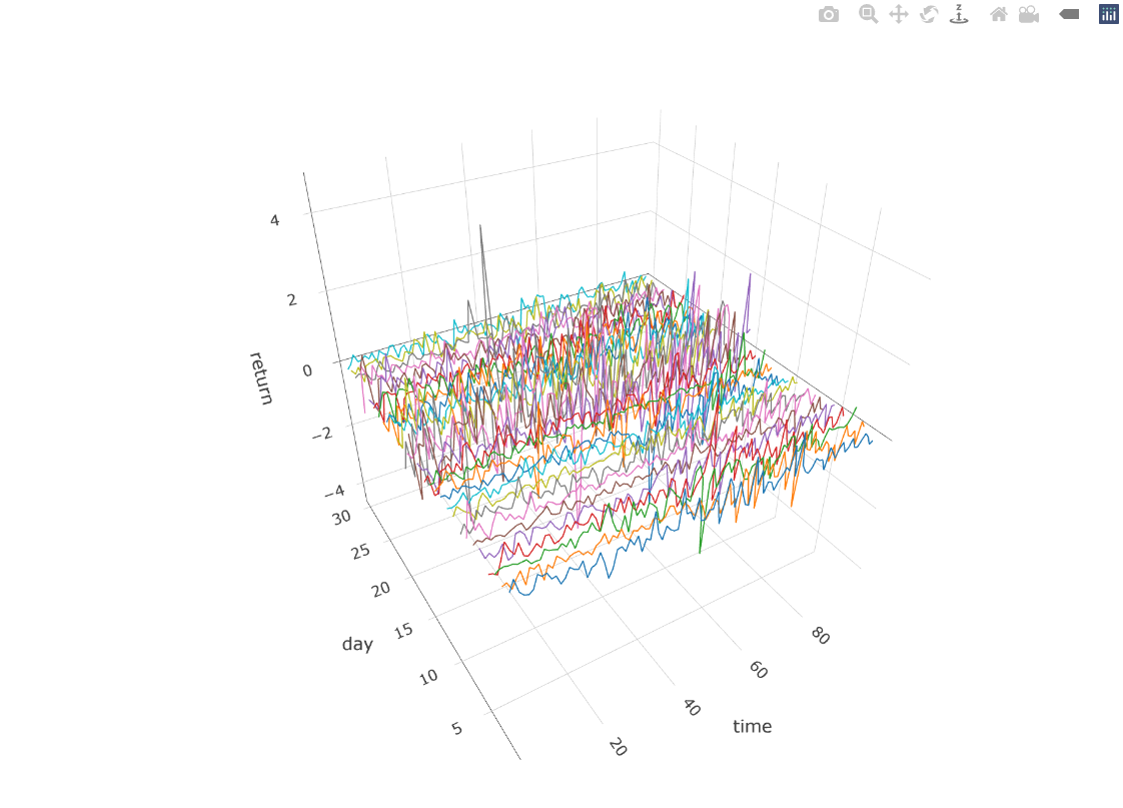}
    			\vspace{0.1cm}
			\hrule
			\vspace{0.1cm}
\caption*{\small \textit {Each line represents one daily function of 15-minute returns}}
    \label{fig: 15min-fun}
\end{figure}

Let us define the 15-minute return on day $i = 1, \dots,N$ as $ X^*_i(t) = (ln(p_t)-ln(p_{t-1})) \times 100$ for $t \in [1,..,96]$ and $N=729$. The functional return series is displayed in Figure \ref{fig: 15min-fun}.

\medskip

\nin  Figure \ref{fig: 15min-fun} illustrates the daily functions of 15-minute BTC returns. Following the approach of \citet{ramsay2005principal}, each function is next demeaned:

\[
X_i (t) = X_i^*(t) - \frac{1}{N} \sum_{i=1}^N X_i(t), \; t=1,...,96
\]

\nin by substracting the 15-minute average return at t=1,...,96 over $N$ days. The demeaned return functions are displayed in Figure \ref{fig:15min-demean} below.

\begin{figure}[H]
    \centering
        \caption{Functions of BTC Demeaned 15-Minute Returns}
    \includegraphics[width=0.5\linewidth]{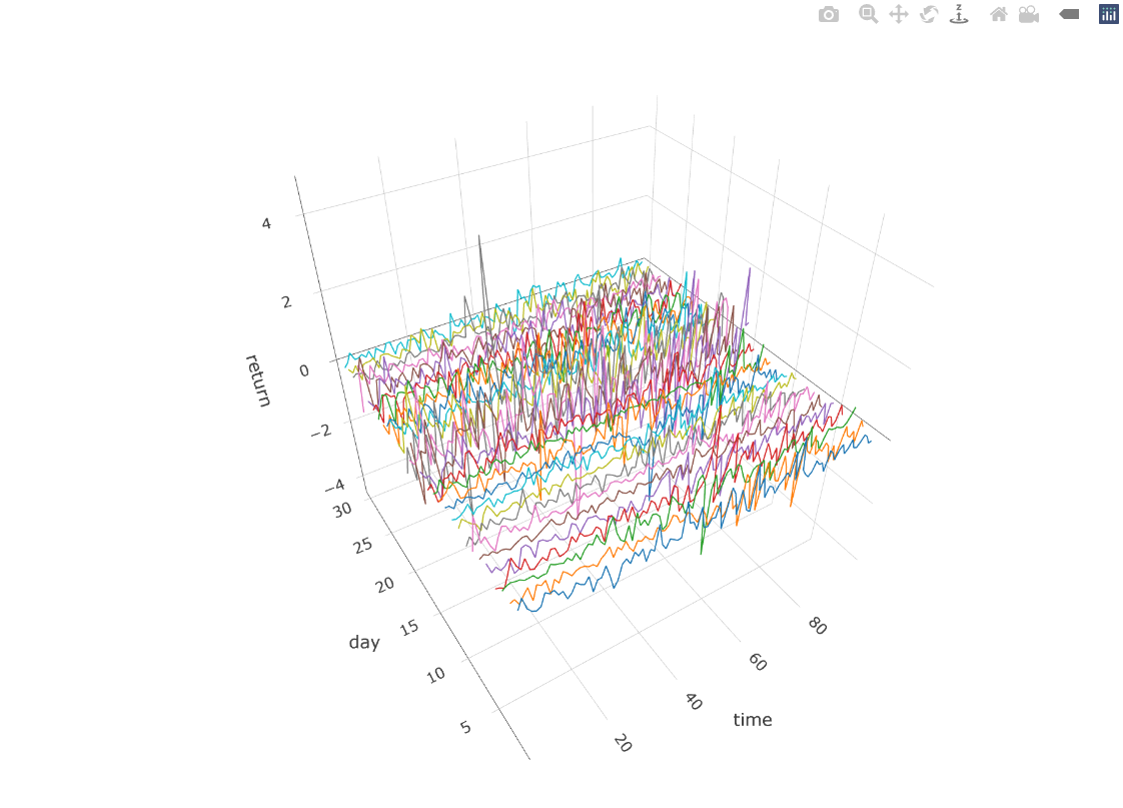}
    			\vspace{0.1cm}
			\hrule
			\vspace{0.1cm}
\caption*{\small \textit {Each line represents one daily demeaned 15-minute return function}}
    \label{fig:15min-demean}
\end{figure}

\nin Figure \ref{fig:15min-demean} shows the daily functions of demeaned 15-minute returns over $N=729$ days.

\subsection{Intraday Hourly Returns}

\begin{figure}[H]
    \centering
        \caption{Functions of BTC Hourly Return}
    \includegraphics[width=0.5\linewidth]{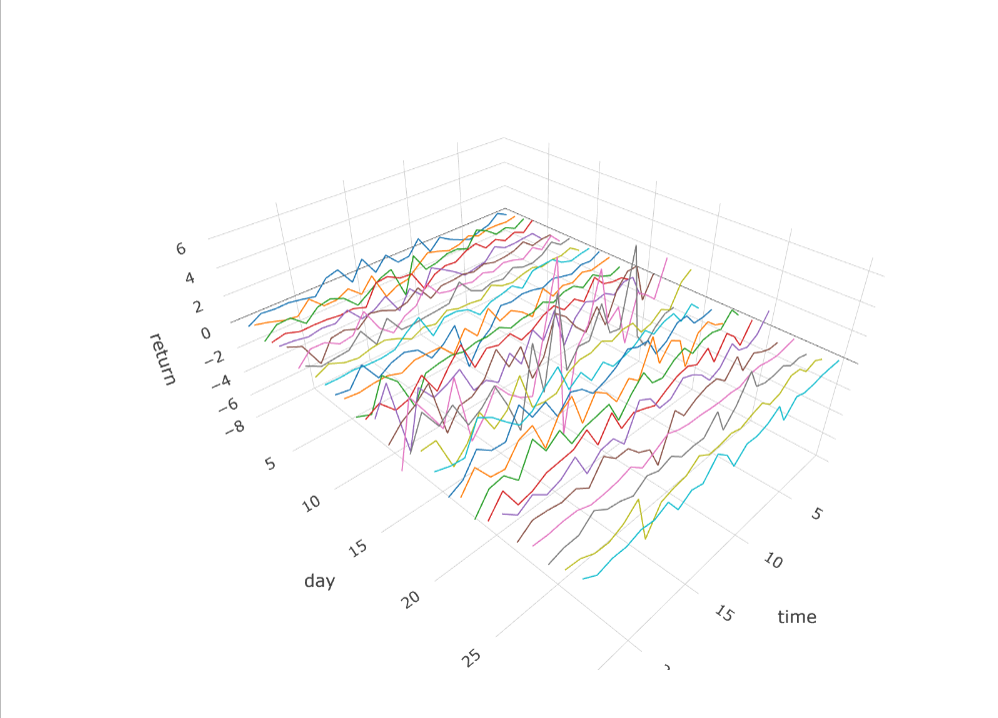}
    			\vspace{0.1cm}
			\hrule
			\vspace{0.1cm}
\caption*{\small \textit {Each line represents a daily return function where returns are sampled hourly}}
    \label{fig:price funs}
\end{figure}

A similar approach is applied to compute the hourly returns from hourly Bitcoin prices over $N=729$ days. The hourly returns are next demeaned.
Figure \ref{fig:price funs}  shows the BTC hourly return functions.

Next, each hourly return function is demeaned:

$$
X_i (t) = X_i^*(t) - \frac{1}{N} \sum_{i=1}^N X_i(t), \; t=1,...,24
$$

\nin by subtracting the hourly average returns at $t=1,...,24$ over $N$ days.
We display the demeaned process in Figure \ref{fig:demean hourly} below:

\begin{figure}[H]
    \centering
        \caption{Functions of BTC Hourly Demeaned Return}
    \includegraphics[width=0.4\linewidth]{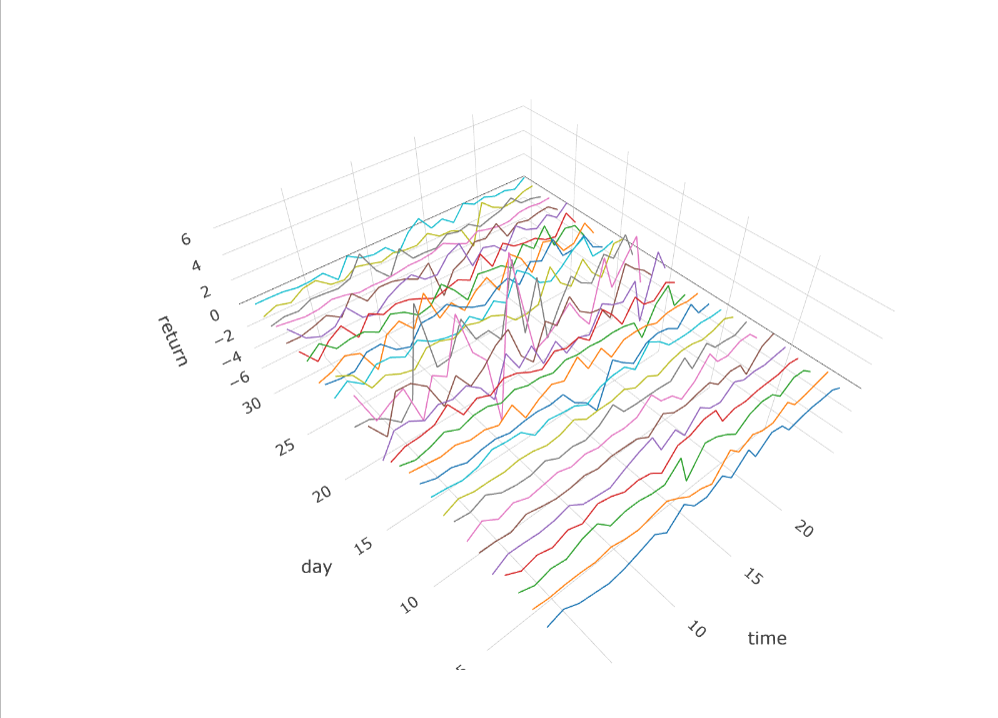}
    			\vspace{0.1cm}
			\hrule
			\vspace{0.1cm}
\caption*{\small \textit {Each line represents one demeaned daily function of hourly returns}}
    \label{fig:demean hourly}
\end{figure}

\subsection{Forecasts of Daily Return Functions}
\medskip
We consider one-day-ahead forecasts of return functions of 15-minute and hourly returns based on $250$ previous daily functions. This approach ensures a sufficient number of in-sample errors allowing us to compare our results with the in-sample error-based forecast intervals of \citet{aue2015prediction}. 

We determine the optimal number $J$ of eigenfunctions by considering the minimum number of leading components that allow us to explain a given level of the cumulative proportion of variance (CPV).  We follow \citep{horvath2012}, who use

\[
J_{\text{CPV}} = argmin_{J: 1 \leq J \leq N} 
\left\{ \frac{\sum\limits_{j=1}^{J} \hat{\lambda}_j}{\sum\limits_{j=1}^{N} \hat{\lambda}_j} \geq \delta \right\},
\]

\nin where $\hat{\lambda}_j$ represents the $J^{th}$ estimated eigenvalue and $\delta = 85 \%$\footnote{In our 15-Minute case, it takes the first 46 (J = 46) eigenvalues to explain 85 \% variance. The number J equals 16 in our hourly case.} \citep{horvath2012}.

Let us illustrate the autocorrelation functions (ACF) and cross ACF for the first three eigenscores in 15-minute data.

\medskip
\begin{figure}[H]
    \centering
        \caption{ACF of 15-Minute Eigenscores Corresponding with Eigenvalues of Order 1 to 3}
    \includegraphics[width=0.9\linewidth]{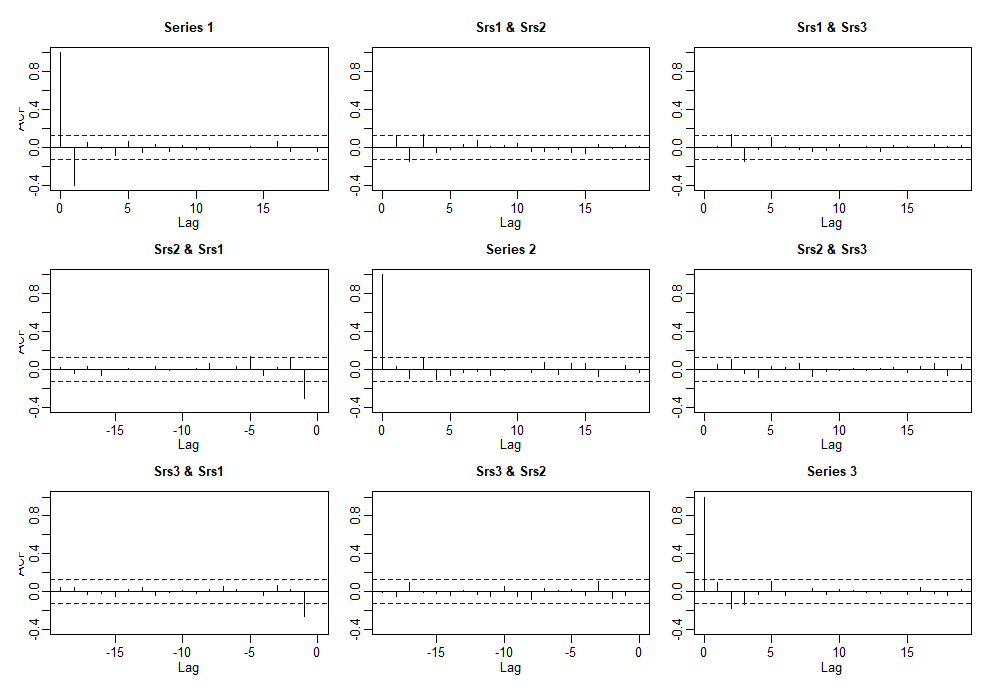}
    			\vspace{0.1cm}
			\hrule
			\vspace{0.1cm}
\caption*{\small \textit {The diagonal graphs show the ACF of each eigenscore $\beta_j, j=1,2,3$. The off-diagonal graphs show the cross ACF between the eigenscores $\beta_j, \beta_k$ for $j\neq k$. }}
    \label{fig:15acf}
\end{figure}

Figure \ref{fig:15acf} shows that the autocorrelations at lag 1 are statistically significant in all $\beta_j, j = 1,\dots, 3$, and in most eigenscores of higher order (not displayed). There also exist statistically significant lagged cross-correlations between $\beta_j, \beta_k$ for $j\neq k$, which motivates the use of the VAR model for the conditional means of eigenscores.  

\begin{figure}[H]
    \centering
        \caption{ACF of 15-Minute Squared Eigenscores Corresponding with Eigenvalues of Order 1 to 3}
    \includegraphics[width=0.9\linewidth]{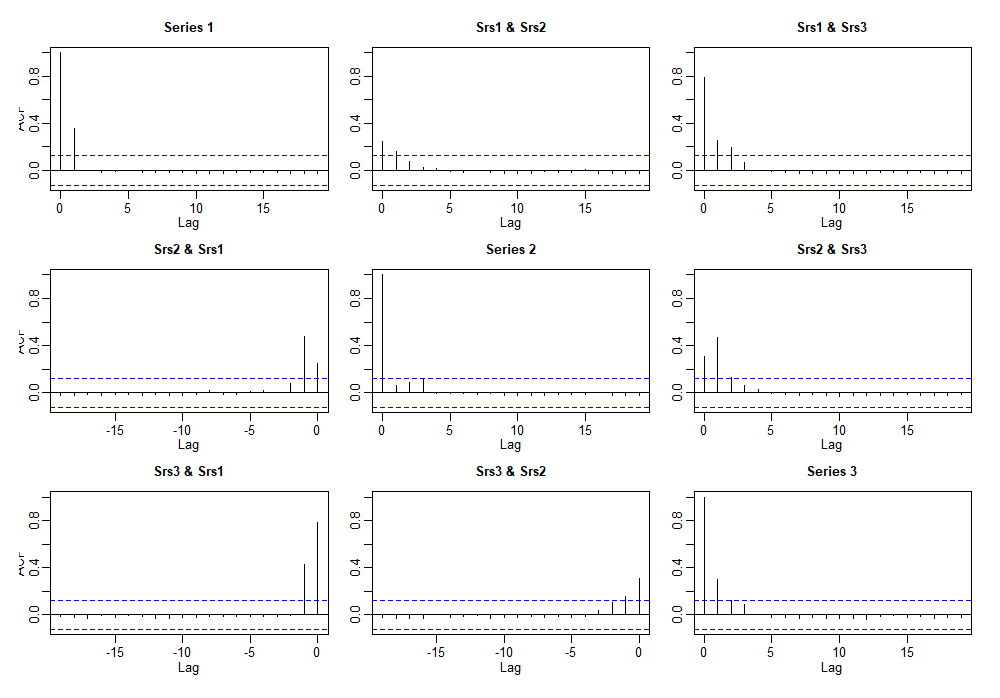}
    			\vspace{0.1cm}
			\hrule
			\vspace{0.1cm}
\caption*{\small \textit {The diagonal graphs show the ACF of the square of each eigenscore $\beta_j, j=1,2,3$. The off-diagonal graphs show the cross ACF between the squares of eigenscores $\beta_j, \beta_k$ for $j\neq k$.}}
    \label{fig:15acf2}
\end{figure}

Figure \ref{fig:15acf2} shows that the GARCH effects exist for $\beta_j, j=1,...3$ and there are also significant laggedd cross-correlations between the squares of $\beta_j, \beta_k$ for $j\neq k$ at short lags. The serial correlation in the eigenscores and their squares motivates us to use the AR-GARCH and VAR-sBEKK models of eigenscores.

The figures displaying the ACF for hourly data can be found in the Appendix.

\subsubsection{Performance Measures}

The root mean square error (RMSE) and mean absolute error (MAE) are used to evaluate the performance of the forecast at discrete points on each future function predicted one-step ahead out of sample. The formulas are: 

\medskip
\nin Root Mean Square Error (RMSE)

\[
RMSE = \sqrt{\frac{1}{n}\sum_{t=1}^n(\hat{x}_t - x_t)^2}
\]

\nin Mean Absolute Error (MAE)
\[
MAE = \frac{1}{n} \sum_{t=1}^n |\hat{x}_t - x_t|
\]

\nin where $n$ is the number of discrete forecast values in a 1-day-ahead function.

To evaluate the interval forecast accuracy, we use the interval score introduced by \citet{GneitingTilmann2007SPSR} and \citet{GneitingTilmann2014PF}.

For pointwise forecast intervals at the $100(1-\omega)\%$ nominal coverage probability, the lower and upper bounds at $\omega/2$ and $1-\omega/2$, denoted by $\hat{X}_{N+1}^{lb}(t)$ and $\hat{X}_{N+1}^{ub}(t)$. The score for the forecast interval is :

\begin{align}
    S_{\omega}[\hat{X}_{N+1}^{lb}(t), \hat{X}_{N+1}^{ub}(t), X_{N+1}(t)] = &[\hat{X}_{N+1}^{ub}(t) - \hat{X}_{N+1}^{lb}(t)] \notag \\  
    &+\frac{2}{\omega}[\hat{X}_{N+1}^{lb}(t)-X_{N+1}(t)]\mathbf{1}\{X_{N+1}(t) < \hat{X}_{N+1}^{lb}(t)\}\notag \\
    &+\frac{2}{\omega}[X_{N+1}(t) - \hat{X}_{N+1}^{ub}(t)]\mathbf{1}\{X_{N+1}(t) > \hat{X}_{N+1}^{ub}(t)\}
    \label{ci}
\end{align}

\nin where $\mathbf{1}\{\cdot\}$ represents the binary indicator function, and $\omega$ is the significance level \citep{GneitingTilmann2007SPSR}. The mean score for the one-day-head forecast interval is :
\[
\bar{S}_\omega = \frac{1}{T}\sum_{t = 1}^T S_{\omega}[\hat{X}_{N+1}^{lb}(t), \hat{X}_{N+1}^{ub}(t), X_{N+1}(t)]
\]

\nin which rewards a narrow forecast interval if, and only if, the true future observation lies within the forecast interval \citep{shang2020dynamic}.

\subsubsection{FPCA Forecast with Univariate AR-GARCH Model of Eigenscores}

In this section, each eigenscore $\beta_j, \; j = 1, \dots, J$ is considered an independent time series that follows the AR(1)-GARCH(1, 1) model to account for the conditional heteroskedasticity.  Our approach is compared with that \citet{shang2022dynamic}, where the eigenscores are modeled as ARMA(p,q) processes\footnote{The orders $p$ and $q$ are automatically selected by the auto.arima package in R.} and the conditional heteroskedasticity is disregarded .

\medskip
\begin{figure}[H]
    \centering
        \caption{One-Day-Ahead 15-Min Return Function Forecast and Forecast Intervals}
    \includegraphics[width=0.7\linewidth]{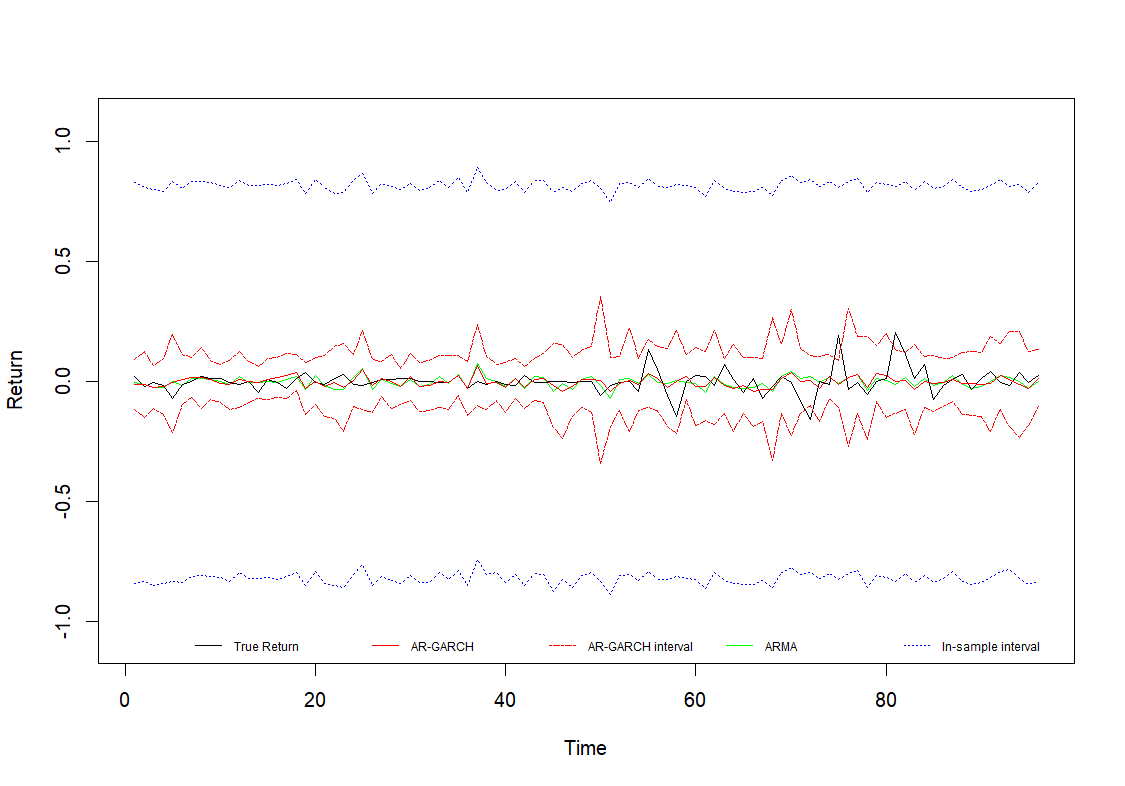}
    			\vspace{0.1cm}
			\hrule
			\vspace{0.1cm}
\caption*{\small \textit {This figure compares the forecast functions and forecast intervals of 15-Min returns on 2022-12-26 based on AR(1)-GARCH(1,1) and ARMA(p,q) models of eigenscores}}
    \label{fig:15argarch}
\end{figure}

\medskip

\begin{figure}[H]
    \centering
        \caption{One-Day-Ahead Hourly Return Function Forecast and Forecast Intervals}
    \includegraphics[width=0.7\linewidth]{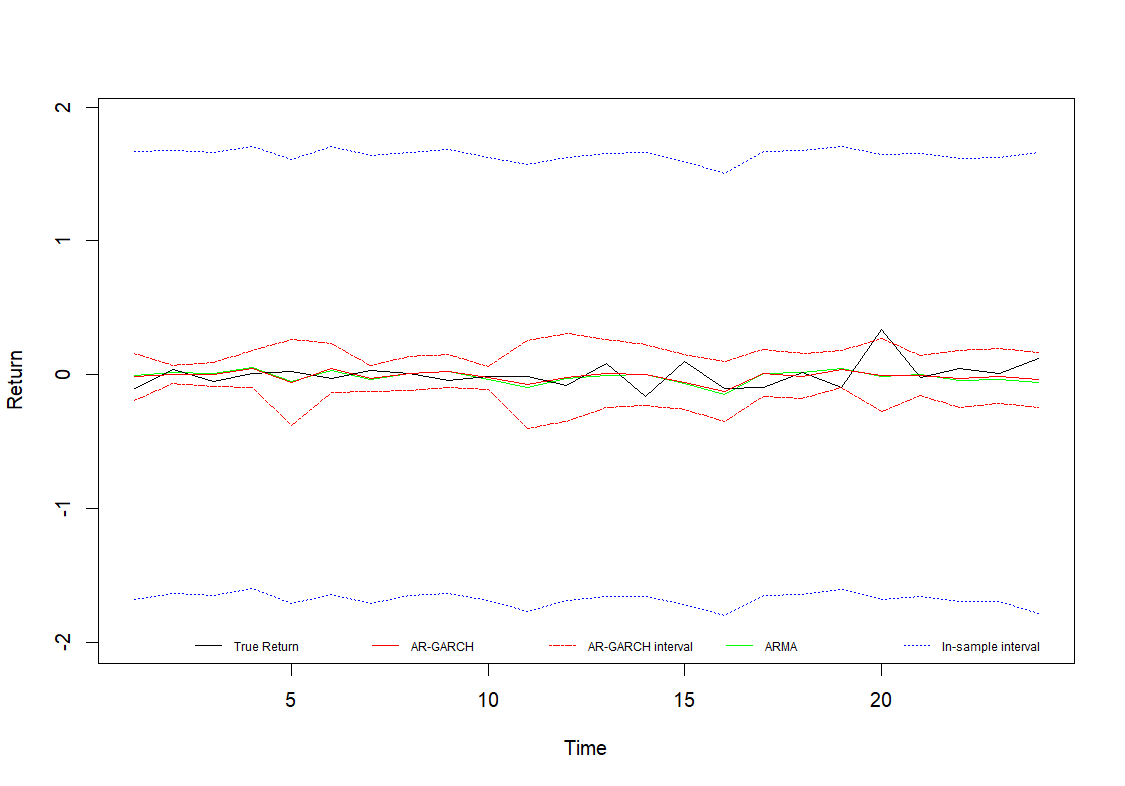}
    			\vspace{0.1cm}
			\hrule
			\vspace{0.1cm}
\caption*{\small \textit {This figure compares the forecast functions and forecast intervals of hourly returns on 2022-12-26 based on AR(1)-GARCH(1,1) and ARMA(p,q) models of eigenscores}}
    \label{fig:hourly argarch}
\end{figure}

\nin Figures \ref{fig:15argarch} and \ref{fig:hourly argarch} show the one-day-ahead forecasts and forecast intervals for the 15-minute and hourly BTC
return functions on 2022-12-26, respectively. The one-day-ahead forecast functions, red for AR(1)-GARCH(1,1) and green for ARMA(p,q), are calculated as one-period ahead conditional mean forecasts of AR-GARCH and ARMA(p,q) models, respectively. 

The AR(1)-GARCH(1,1) forecast intervals of eigenscores are calculated using formula \ref{fun: garchci} in Section (2.2). The ARMA(p,q) forecast intervals are calculated following \citet{aue2015prediction} and described in Section (2.1). 

\medskip

We observe that our forecast function overlaps the forecast function of \citet{shang2020dynamic} and \citet{shang2022dynamic} and they are both close to the true functions of returns on Bitcoin. 
We also observe a difference in the forecast intervals as those obtained from the AR(1)-GARCH(1,1) model are narrower than those of \citet{aue2015prediction}. This concerns both return series sampled at 15-minute and hourly intervals in Figures \ref{fig:15argarch} and \ref{fig:hourly argarch}.

\begin{table}[H]
    \centering
        \caption{ Forecast Performance: AR(1)-GARCH(1,1) and AR(p,q) Eigenscore Models}
    \begin{tabular}{lcccc}
        \toprule
        \textbf{15-Min} & \textbf{RMSE} & \textbf{MAE} & \textbf{Sign} &\textbf{$\bar{S}_{\omega}$} \\
        \midrule
        \textbf{AR-GARCH} & \textbf{0.087} & \textbf{0.056} & \textbf{47.9 \%} & \textbf{0.607}\\
        ARMA & 0.086 & 0.056 & 46.7\% & 1.374\\
        \bottomrule
        \addlinespace  
        \toprule
        \textbf{Hourly} & \textbf{RMSE} & \textbf{MAE} & \textbf{Sign} &\textbf{$\bar{S}_{\omega}$} \\
        \toprule
        \textbf{AR-GARCH} & \textbf{0.168} & \textbf{0.117} & \textbf{49.7 \%} & \textbf{1.323}\\
        ARMA & 0.168 & 0.117 & 48.2 & 3.672\\
        \bottomrule
    \end{tabular}
    			\vspace{0.1cm}
			\hrule
			\vspace{0.1cm}
\caption*{\small \textit {The results in this table are calculated from 960 discrete forecast 15-minute returns and 240 discrete forecast hourly returns, starting from 2022-12-26.}}
    \label{tab:argarch}
\end{table}

Table \ref{tab:argarch} shows the performance statistics of one-day-ahead return function forecasts over 10 future days starting from 2022-12-26 computed from the AR(1)-GARCH(1,1) and ARMA(p,q) models of eigenscores. Columns 1 to 5 report the time series model of eigenscores (Col.1),
the RMSE (Col.2), MAE (Col.3), the percentage of correctly forecast returns (Col.4) and the interval score (Col.5).
The accuracy of forecast return functions evaluated at discrete points from both models is close. We applied the Diebold-Mariano test to the forecast errors of return functions at discrete points \(e_t = y_t - \hat{y}_t\), and based on the ARMA(p,q) models and AR(1)-GARCH(1,1) models of eigenscores. Both hourly (p-value of 0.362) and 15-minute (p-value of 0.201) results do not reject the null hypothesis, which suggests that the models have the same forecast accuracy level. 

The difference is in the accuracy of pointwise forecast intervals.
The results indicate that the forecast interval estimated by the AR(1)-GARCH(1,1) model is narrower than the forecast interval of \citet{aue2015prediction} based on ARMA(p,q) as in \citep{shang2022dynamic} and \citep{shang2020dynamic} for both 15-minute and hourly return functions.  Our method also outperforms the competitors in terms of the correct sign predictions, reported in Column 4 as "Sign", and of the score of interval predictions \textbf{$\bar{S}_{\omega}$} (the lower, the better), given in Column 5. We present the forecast interval corresponding to a pointwise nominal 95\% level ($\omega = 0.05$ in equation (\ref{ci})). The actual coverage rate achieved by our method, indicating the proportion of times true future returns fall within the forecast intervals, is 98.33\% for the 15-minute returns and 96.35\% for the hourly returns.

\bigskip
\subsubsection{FPCA Forecast with VAR-sBEKK Model of Eigenscores}

\begin{figure}[H]
    \centering
        \caption{One-Day-Ahead 15-Min Return Function Forecast and Forecast Interval (VAR-sBEKK)}
    \includegraphics[width=0.9\linewidth]{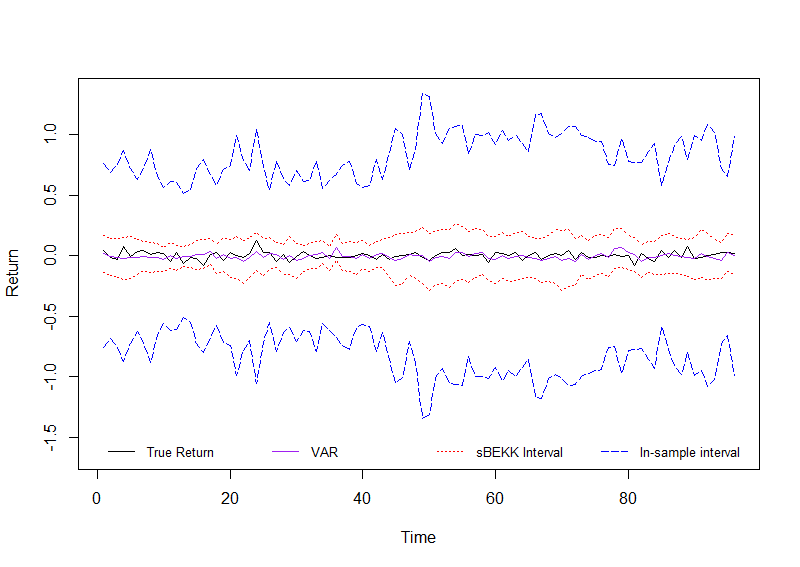}
    			\vspace{0.1cm}
			\hrule
			\vspace{0.1cm}
\caption*{\small \textit {This figure compares the forecast function and forecast interval of 15-Min returns based on the VAR-sBEKK and VAR models of eigenscores}}
    \label{fig:15bekk}
\end{figure}

\begin{figure}[H]
    \centering
        \caption{One-Day-Ahead 15-Min Return Function Forecast and Forecast Interval (VAR-sBEKK)}
    \includegraphics[width=0.9\linewidth]{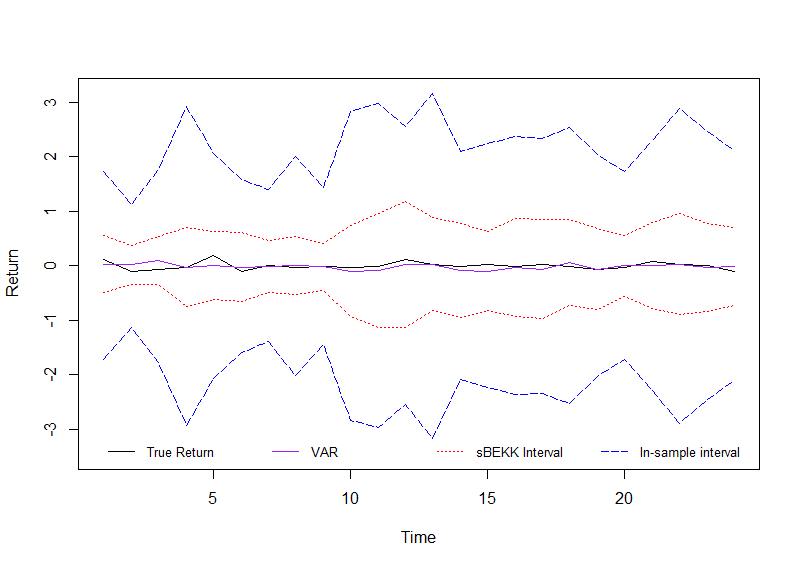}
    			\vspace{0.1cm}
			\hrule
			\vspace{0.1cm}
\caption*{\small \textit {This figure compares the forecast function and forecast interval of hourly returns based on the VAR-sBEKK and VAR models of eigenscores}}
    \label{fig:bekk hour}
\end{figure}

In this section, we take into account the cross-correlation and conditional heteroskedasticity of eigenscores by using the VAR(1)-sBEKK(1,1) model (equation \ref{equ: bekk}).  Figures \ref{fig:15bekk} and \ref{fig:bekk hour} compare the forecast function and forecast interval of 15-minute and hourly returns based on the VAR(1)-sBEKK(1,1) model (11) of eigenscores with the forecast function and forecast interval obtained from the VAR(1) model following the approach of  \citet{aue2015prediction}.

\medskip

\medskip

\medskip

We observe again that the proposed method based on the VAR(1)-sBEKK(1,1) model of eigenscores produces forecast intervals that are narrower than those proposed by \citet{aue2015prediction}.

\begin{table}[H]
    \centering
        \caption{ Forecast Performance: VAR-sBEKK and VAR Eigenscore Models}
    \begin{tabular}{lccccc}
        \toprule
        \textbf{15-Min} & \textbf{RMSE} & \textbf{MAE} & \textbf{Sign} & $\mathbf{\bar{S}_{BEKK}}$ & $\mathbf{\bar{S}_{Aue}}$ \\
        \midrule
         & \textbf{0.089} & \textbf{0.059} & \textbf{47.6 \%} & \textbf{0.552} & \textbf{1.567} \\

        \bottomrule
        \addlinespace  
        \toprule
        \textbf{Hourly} & \textbf{RMSE} & \textbf{MAE} & \textbf{Sign} & $\mathbf{\bar{S}_{BEKK}}$ & $\mathbf{\bar{S}_{Aue}}$ \\
        \toprule
         & \textbf{0.185} & \textbf{0.132} & \textbf{45.0\%} & \textbf{1.443}  &  \textbf{4.115}  \\
        \bottomrule
    \end{tabular}
        			\vspace{0.1cm}
			\hrule
			\vspace{0.1cm}
\caption*{\small \textit {The results in this table are calculated based on the forecast of the same time interval as in Table 1. $\mathbf{ \bar{S}_{BEKK}}$ represents the forecast interval score of the VAR(1)- sBEKK(1,1) model and $\mathbf{\bar{S}_{Aue}}$ represents the forecast interval score of \citet{aue2015prediction}.  }}
    \label{tab:bekk}
\end{table}

Table \ref{tab:bekk} describes the performance of pointwise forecasts and interval forecasts of the same time interval as Section 3.3.2, obtained by applying VAR(1) and VAR(1)-sBEKK(1,1) \footnote{The function \textit{"bekk\_fit"} from the \textit{BEKKs R} package became unstable for a large number of variables includes. The functional variables of 15-minute returns are smoothed using a relatively smaller number of basis functions, so the number of eigenfunctions $J$ included in the sBEKK is small enough to keep \textit{"bekk\_fit"} stable. }
Columns 1 to 5 report the observed frequency, followed by the RMS (Col.2), MAE (Col.3), the percentage of correct forecast sign  (Col.4) and the forecast interval score (Col.5). The RMSE and MAE for our method are identical to those based on the VAR(1) model of \citet{aue2015prediction}, because our estimation proceeds in two steps, and the first step produces the conditional mean forecasts. The \textbf{Sign} in Column 4 provides the percentage of successful sign forecast, which are also identical. The advantage of our approach is in the accuracy of the pointwise forecast interval, which can be evaluated by comparing the interval score $\mathbf{S_{BEKK}}$ of our model to the score $\mathbf{S_{AUE}}$ in Column 5 based on the VAR forecast interval of \citet{aue2015prediction}. The results indicate that the VAR-sBEKK model outperforms the VAR-based forecast of \citet{aue2015prediction} in terms of the accuracy of the forecast intervals in both 15-minute and hourly returns. For the 15-minute returns case, our method's coverage rate, corresponding to the nominal 95\% confidence interval ($\omega = 0.05$ in equation (\ref{ci})), is 97.91\%, whereas for the hourly returns case, it stands at 98.75\%.

By comparing the results in Tables 1 and 2, we find that the multivariate approach based on the VAR-sBEKK model performs the best in terms of the interval accuracy score in the 15-minute return case, while the univariate AR-GARCH model performs the best in terms of the interval accuracy score in the hourly returns.

\section{Intraday FPCA Forecast }

\medskip
So far, we have considered predicting the entire function of returns out-of-sample one day ahead, i.e. for $i=N+1$. Let us now consider a new approach for forecasting returns at shorter intraday horizons of $k$ time units, over 1 or more hours on a given day $i=N$ when the returns are observed only up to a given hour on that day. For assets traded 24/7, the day is conventionally determined by the UTC hours 0:00 and 24:00. A daily function of returns may as well start at a different point and cover 24 consecutive hours. This idea underlies the forecasting approach based on a "rolling FPCA" and inspired by the paper of \citet{aguilera1997approximated}.

\subsection{FPCA Rolling Forecast Model}

Let us consider the following scenario: On day $N$, We observe $N-1$ complete functions $X_i(t)$ on the interval \(t=[1, \dots, 24], \; i = 1,\dots, N-1\). Suppose that on day $N$ the return function is incomplete: we have observed only the returns up to and including t=23, and we wish to predict the return over the last hour, i.e in the neighborhood of $t=24$. To do that, we consider an auxiliary set of complete functions  $X_i(s)^+$ on the interval $s=[0, \dots, 23]$, as illustrated in Table 3 below, observed on days $i=1,..., N$.

First, we perform the FPCA on the set of $N-1$ functions to get the eigenscores \(\hat{\beta}_{ij}, \; i = 1,\dots, N-1, \;j = 1,\dots,J\) and eigenfunctions $\hat{\xi}_l^*(t)$.

Next, we perform the FPCA on the set of $N$ auxiliary functions $X_i(s)^+, s=[0,\dots,23], \; i = 1,\dots,N$, yielding eigenscores $\hat{\alpha}_{ij}, i = 1,\dots,N, \;j = 1,\dots,J$, and eigenfunctions $\hat{\xi}_j^+(s)$. In order to predict the neighborhood of $X_N(24)$, we need to predict the eigenscores $\beta_{ij}$ associated with $X_N(t =1, \dots,24)$. This is done by a regression model.  

\medskip
\begin{table}[H]
  \centering
    \caption{Illustration of rolling functions of time}
  \resizebox{\textwidth}{!}{\begin{tabular}{|c|c|c|c|c|c|c|c|c|c|c|c|c|c|c|c|c|c|c|c|c|c|c|c|c|c|}
    \hline
Hour & 0(24) & 1 & 2 & 3 & 4 & 5 & 6 & 7 & 8 & 9 & 10 & 11 & 12 & 13 & 14 & 15 & 16 & 17 & 18 & 19 & 20 & 21 & 22 & 23 & 24  \\ \hline
  $X^+(s)$& \textcolor{blue}{\textbullet{}} & \textcolor{blue}{\textbullet{}} & \textcolor{blue}{\textbullet{}} & \textcolor{blue}{\textbullet{}} & \textcolor{blue}{\textbullet{}} & \textcolor{blue}{\textbullet{}} & \textcolor{blue}{\textbullet{}} & \textcolor{blue}{\textbullet{}} & \textcolor{blue}{\textbullet{}} & \textcolor{blue}{\textbullet{}} & \textcolor{blue}{\textbullet{}} & \textcolor{blue}{\textbullet{}} & \textcolor{blue}{\textbullet{}} & \textcolor{blue}{\textbullet{}} & \textcolor{blue}{\textbullet{}} & \textcolor{blue}{\textbullet{}} & \textcolor{blue}{\textbullet{}} & \textcolor{blue}{\textbullet{}} & \textcolor{blue}{\textbullet{}} & \textcolor{blue}{\textbullet{}} & \textcolor{blue}{\textbullet{}} & \textcolor{blue}{\textbullet{}} & \textcolor{blue}{\textbullet{}} &\textcolor{blue}{\textbullet{}} & \\ \hline
  $X(t)$& & \textcolor{red}{\textbullet{}} & \textcolor{red}{\textbullet{}} & \textcolor{red}{\textbullet{}} & \textcolor{red}{\textbullet{}} & \textcolor{red}{\textbullet{}} & \textcolor{red}{\textbullet{}} & \textcolor{red}{\textbullet{}} & \textcolor{red}{\textbullet{}} & \textcolor{red}{\textbullet{}} & \textcolor{red}{\textbullet{}} & \textcolor{red}{\textbullet{}} & \textcolor{red}{\textbullet{}} & \textcolor{red}{\textbullet{}} & \textcolor{red}{\textbullet{}} & \textcolor{red}{\textbullet{}} & \textcolor{red}{\textbullet{}} & \textcolor{red}{\textbullet{}} & \textcolor{red}{\textbullet{}} & \textcolor{red}{\textbullet{}} & \textcolor{red}{\textbullet{}} & \textcolor{red}{\textbullet{}} & \textcolor{red}{\textbullet{}} & \textcolor{red}{\textbullet{}} & \textcolor{red}{\textbullet{}} \\ \hline
  \end{tabular}}

      			\vspace{0.3cm}
			\hrule
			\vspace{0.1cm}
\caption*{\small \textit {$X^+(s)$ and $X(t)$} are both functions on 24 hourly returns with $X^+(s)$ shifted one-hour-behind of $X(t)$}
  \label{tab:hour_graph}
\end{table}
\medskip

\nin Table \ref{tab:hour_graph} provides an example of $X^+(s)$ and $X(t)$ when forecasting the return function over hour $24$ of UTC time. Note that there are 23 overlapping discrete hourly returns between the functions $X^+(s)$ and $X(t)$. For a prediction at horizon $k>1$, there will be $24-k$ overlapping discrete returns. The algorithm proceeds as follows:

\medskip
\nin \textbf{Step 1}: Calculate $N$ demeaned functions $X_{i}^+(s)$ by subtracting the sample mean $\bar{X}^+(s) = \frac{1}{N}\sum_{i = 1}^N X_{i}^+(s), \; s = 1-k,\dots, 24-k$. 

\nin \textbf{Step 2}: Calculate $N-1$ demeaned functions $X_{i}(t)$ by subtracting the sample mean $\bar{X}(t) = \frac{1}{N-1}\sum_{i = 1}^N X_{i}(t), \; t = 1,..., 24$.

\nin \textbf{Step 3}: Estimate the eigenfunctions $\hat{\xi}_{j}^+(s), j = 1,\dots, J$  from the covariance operator of $X_{i}^+(s), i = 1, \dots, N$  and compute eigenscores $\hat{\alpha}_{ij} = <X_{i}^+, \hat{\xi}_{j}^+(s)>, i = 1,\dots, N; \; j = 1,\dots, J$  of $X^+(s)$ by FPCA. 

\nin \textbf{Step 4}: Estimate the eigenfunctions $\hat{\xi}_{l}(t), l = 1, \dots, J$ from the covariance operator of $X_{i}(t), i = 1, \dots, N-1$, and compute the eigenscores $\hat{\beta}_{il} = <X_{i}, \hat{\xi}_{l}(t)>$ by FPCA. 


\nin \textbf{Step 5}: For each $j=1,..,J$ run a regression model of a vector of  eigenscores $\hat{\beta}_{.j}= [\beta_{1,j},...,\beta_{N-1,J}]'$ of length $N-1$  on the  $(N-1) \times J$ matrix $\hat{\alpha}= [\alpha_{1,l},...,\alpha_{N-1,J}]'$, $i=1,...,N-1$:

\[
\hat{\beta}_{.j} = \hat{\alpha} c + v_j,  \;\; \mbox{for} \; j = 1, \dots, J ,
\]
\nin where $c = [c_1,..,c_J]'$ is a vector of coefficients and $v_j$ is the regression error with mean 0 and finite variance. 

\nin \textbf{Step 6}: Predict the eigenscores $\beta_{N, j}, \; j = 1, \dots, J$ of length $J$ on day $N$ by using the estimated regression coefficients from \textbf{Step 4} and the $1 \times J$ row vector of eigenscores $\hat{\alpha}_{N}$ estimated in \textbf{Step 3} as the explanatory variable:

\[
\tilde{\beta}_{N,j} =  \hat{\alpha}_{N} c  \;\; \mbox{for} \; j = 1, \dots, J ,
\]

\nin \textbf{Step 7}: The forecast over the 24th hour on the $N^{th}$ function is obtained from:

$\hat{X}_{N}(t) = \sum_{j = 1}^J \tilde{\beta}_{Nj} \hat{\xi}_{j}(t) + \bar{X}(t). $

\nin The forecast function $\hat{X}_{N}(24)$ evaluated at $t=24$, gives the value  is the forecast value of the one-hour-ahead target return. 

\medskip
Note that the FPCA performed on $X(t)$ and $X^+(s)$ can differ
because the stationarity is satisfied over days $i=1,..N$. Hence, the covariances and their principal components can vary over the subsets of time $t$, although this variation is bounded by the square integrability of the functions. 

\subsection{FPCA Rolling Forecast Model: Application to Bitcoin Returns}

We forecast the hourly returns on Bitcoin out-of-sample at horizon $k=1$ of one hour for a continuous 200 discrete hourly returns starting from 2019-02-07, which is the same out-of-sample period as \citet{gradojevic2023forecasting}'s subsample 5 for hourly BTC returns. 
There is no optimal way to determine how to choose the best number of daily functions to initiate the algorithm. We repeat the above process for a chosen range of daily function numbers of $[90,110]$. The results indicate that applying different numbers of daily functions only changes the MSE slightly; however, the number of daily functions applied does affect the correct forecast sign rate. Therefore, in this paper, the number of daily functions applied to the algorithm is chosen by the value that results in the best correct forecast sign rate. 

\medskip
\begin{figure}[H]
    \centering
        \caption{Rolling FPCA: Hourly BTC Return Forecast}
    \includegraphics[width=0.9\linewidth]{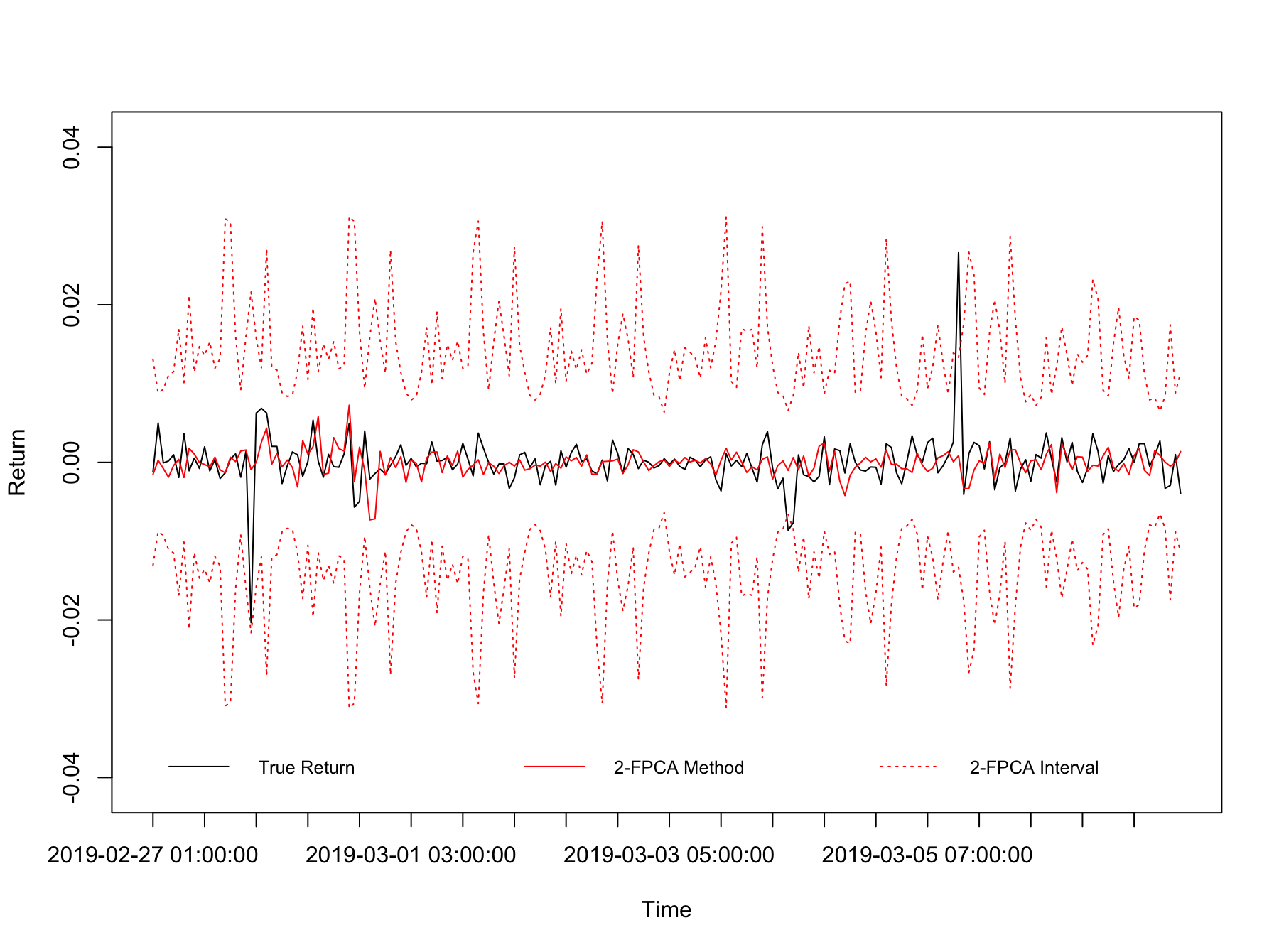}
    			\vspace{0.1cm}
			\hrule
			\vspace{0.1cm}
\caption*{\small \textit { }}
    \label{fig:2fpca}
\end{figure}

\medskip
In Step 5 we use various estimators of coefficients $c$. Because we expect the off-diagonal elements of the  $\alpha'\alpha$ matrix to be close to zero by the orthogonality conditions, and the diagonal elements for large $j$ to be close to zero as well, we consider the Lasso and Ridge estimators, to ensure that this matrix is non-singular. 
The Support Vector Machine (SVM) is used as it provides optimal weighting and bias correction.
We also account for potential non-linear relations between the eigenscores by using the  Random forest (RF), and Neural Networks (NN).

We compare our forecast performance with the out-of-sample forecast for subsample 5 of \citet{gradojevic2023forecasting}, who used the Feedforward Deep Artificial Neural Network(FF-D-ANN), SVM, RF estimators of the Random Walk (RW) and ARMAX(1,1) (Autoregressive Moving Average with exogenous inputs) models to forecast one-hour-ahead BTC returns. 

\begin{table}[H]
    \centering
        \caption{Forecast Performance with Different Regression Estimators in Step 5}
    \begin{tabular}{|c|c|c|} \hline 
         Estimator &  RMSE & Sign (\%)\\ \hline 
         OLS&  0.0345& 59.0\\ \hline 
         Ridge OLS&  0.0331& 62.5\\ \hline 
         LASSO OLS&  0.0342& 63\\ \hline 
         SVM&  0.0321& 57.5\\ \hline 
         RF&  0.0335& 60.5\\ \hline 
         NN&  0.0437& 53.5\\ \hline
    \end{tabular}
        			\vspace{0.1cm}
			\hrule
			\vspace{0.1cm}
\caption*{\small \textit {The results in this table are calculated by 200 discrete forecast hourly returns starting from 2019-02-07. }}
    \label{tab:my_label}
\end{table}
Our proposed algorithm gives the best performance in terms of forecast error when the SVM method is used in \textbf{Step 5}, which yields an RMSE of $0.0321$ while the Ridge OLS results in the best forecast sign rate of $62.5\%$. In \citet{gradojevic2023forecasting}, the Random Walk model performs best, resulting in an RMSE of $0.0363$ and a correct forecast sign rate of $50.24\%$.

\subsection{FPCA Rolling Forecast Model: the Horizon Effect}

So far we have discussed the forecasts at horizon $k=1$. At $K=1$, the coefficients $\alpha$ are arbitrarily close to $\beta$, so that the coefficient of determination $R^2$ of the Ridge, or Lasso regression in step 5 is arbitrarily close to 1.
When the forecast horizon $k$ increases, the auxiliary set of functions $X^+(s)$ gets shifted backward in time by $k$ and the overlap between $X(t)$ and $X^+(s)$ diminishes. Then, the explanatory power of coefficients $\alpha$ decreases with the horizon $k$ and the $R^2$ of the regression in step 5 diminishes quickly.  We observe that the $R^2$ is above 0.5 up to about horizon $k=10$, as illustrated in Figure 12 below.


\begin{figure}[H]
    \centering
        \caption{$R^2$ for Different Horizons of Using LASSO In Step 5 }
    \includegraphics[width=0.4\linewidth]{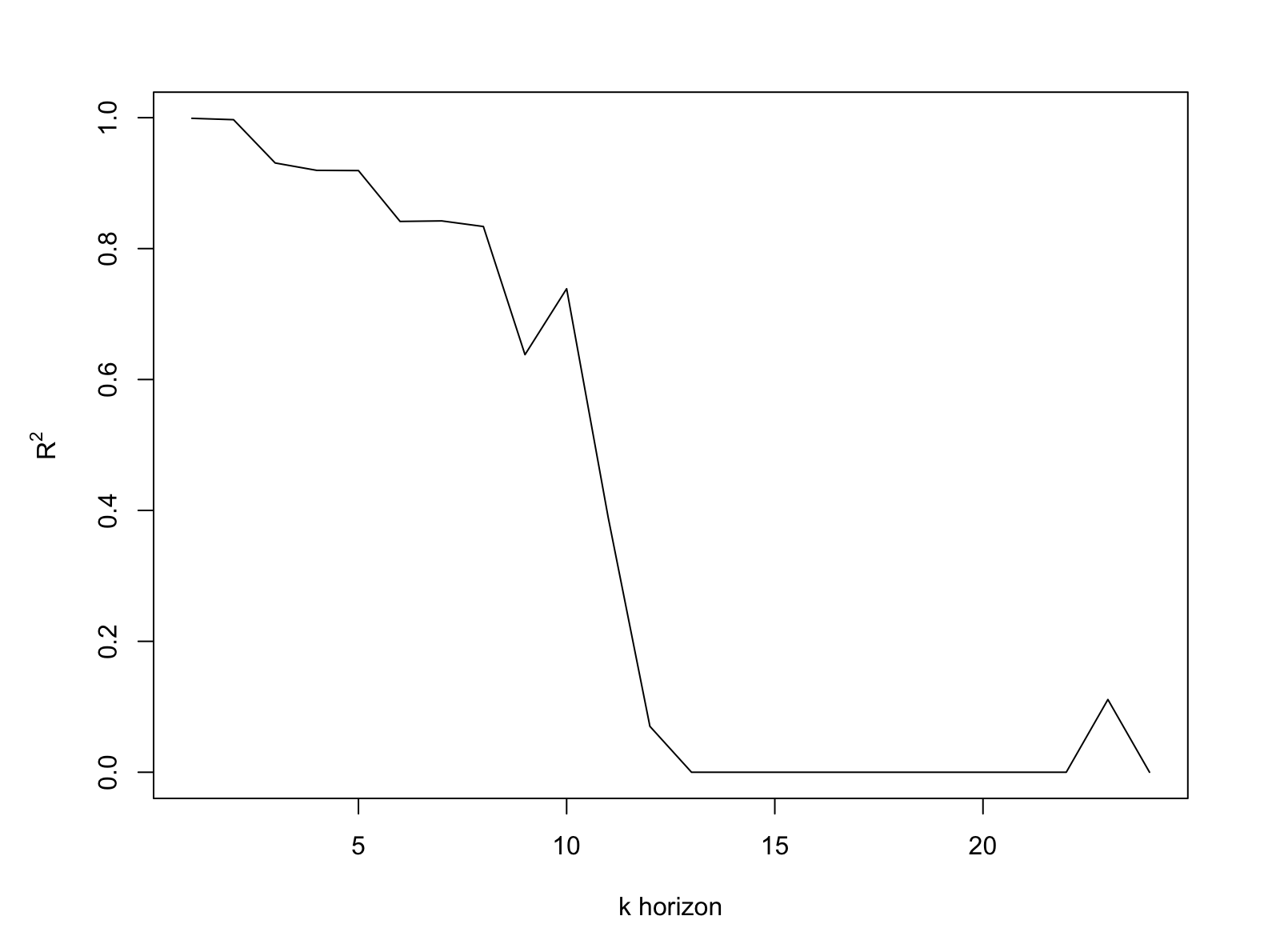}
    			\vspace{0.1cm}
			\hrule
			\vspace{0.1cm}
\caption*{\small \textit { }}
    \label{fig:kmes}
\end{figure}

\nin Hence we recommend using the above intraday forecasting method at very short horizons.

\section{Conclusions}
\medskip

In this paper, we define the KL dynamic factor model for the analysis of conditionally heteroscedastic functional processes. We also introduce two new Functional Principal Component Analysis (FPCA) based methods for forecasting intraday Bitcoin return. The first approach aims at improved interval forecasting of daily returns and exploits the serial correlation of eigenscores revealed in the FPCA literature. 
The novelty of our approach is that it also takes into consideration the conditional heteroscedasticity of returns and produces pointwise forecast intervals that are narrower than the existing literature.
The second approach is a rolling FPCA for intraday forecasting of Bitcoin returns. It allows us to consider a sequence of partially overlapping daily return functions that start at subsequent time points and forecast the returns intradaily. 

The empirical results of forecasting Bitcoin daily 15-minute and hourly  return functions from the proposed methods indicate that accounting for the conditional heteroscedasticity of returns by forecasting the eigenscores using AR(1)-GARCH(1,1) and VAR(1)-sBEKK(1,1) models noticeably improves the accuracy of forecast intervals. Our proposed "rolling" FPCA method for forecasting one-step-ahead hourly Bitcoin returns shows better performance in terms of both forecast errors and forecast sign accuracy compared to other methods of Bitcoin forecasting in the literature \citep{gradojevic2023forecasting}.

\newpage

\nocite{KokoszkaP.2015FDFM, horman2013functional}
\bibliographystyle{apalike}
\bibliography{references}

\newpage
\section*{Appendix A}

\begin{figure}[H]
    \centering
        \caption{ACF of Hourly Eigenscores Corresponding with Eigenvalue Order 1 to 3}
    \includegraphics[width=0.9\linewidth]{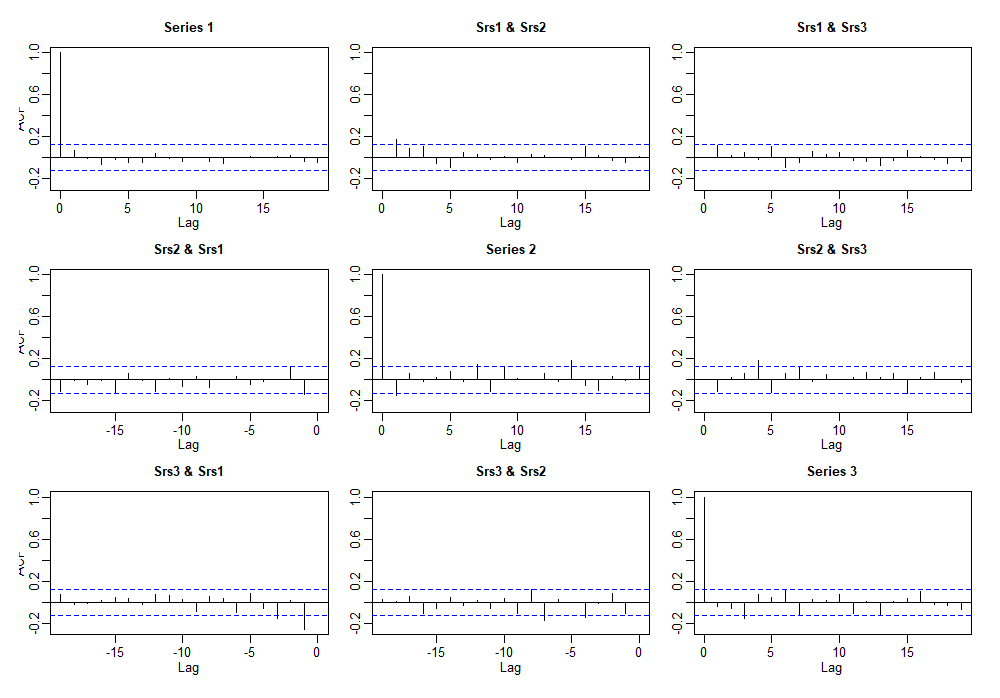}
    			\vspace{0.1cm}
			\hrule
			\vspace{0.1cm}
\caption*{\small \textit {The diagonal graphs show the ACF of each eigenscore. The off-diagonal graphs show the cross ACF between different eigenscores. }}
    \label{fig:hourlyacf}
\end{figure}

\begin{figure}[H]
    \centering
        \caption{ACF of Hourly Square of Eigenscores Corresponding with Eigenvalue Order 1 to 3}
    \includegraphics[width=0.9\linewidth]{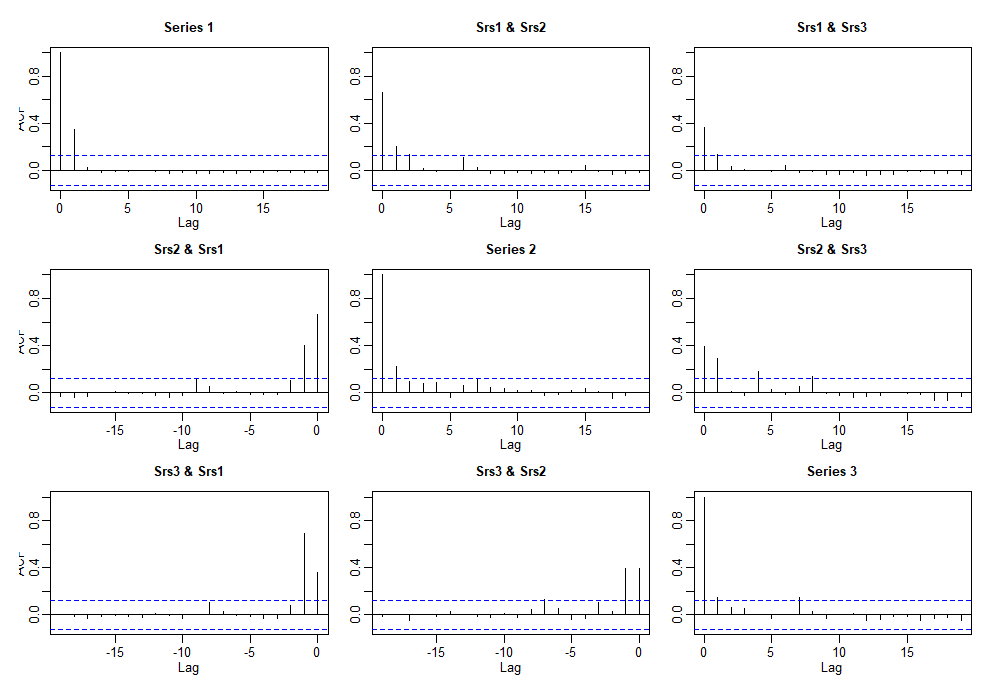}
    			\vspace{0.1cm}
			\hrule
			\vspace{0.1cm}
\caption*{\small \textit {The diagonal graphs show the ACF of the square of each eigenscore. The off-diagonal graphs show the cross ACF between different squares of eigenscores.}}
    \label{fig:hourly acf2}
\end{figure}


\begin{figure}[H]
    \centering
        \caption{Daily Variance of 15-Minute BTC Return and Daily Variance of In-sample Estimation Errors}
    \includegraphics[width=0.9\linewidth]{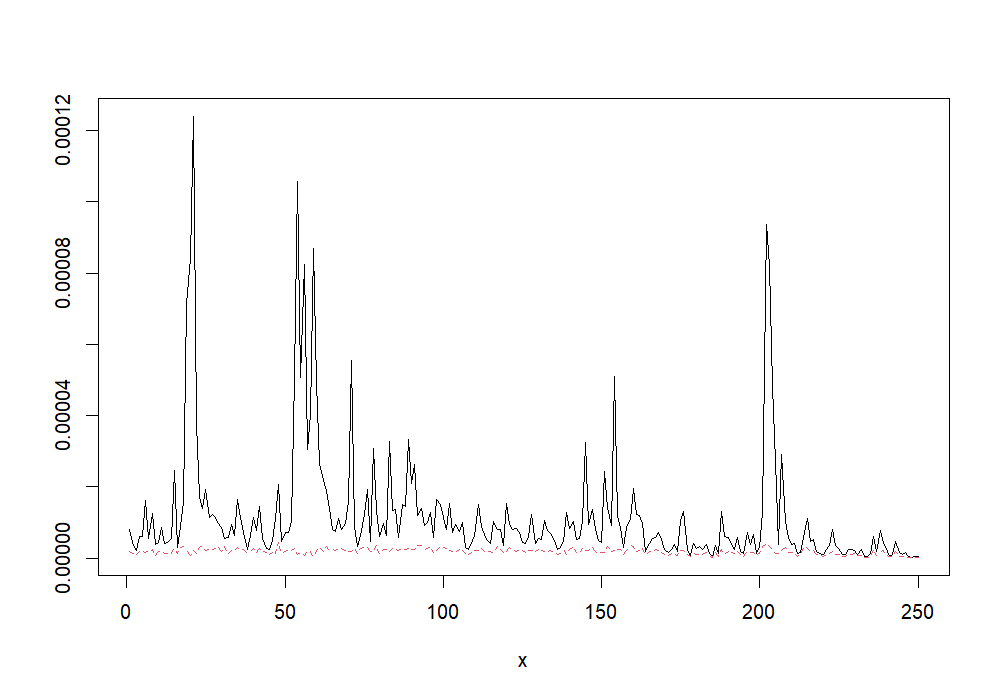}
    			\vspace{0.1cm}
			\hrule
			\vspace{0.1cm}
\caption*{\small \textit {The black line illustrates the daily variance of 15-Minute BTC returns. The red line illustrates the daily variance of the in-sample error $\hat{e}_i(t) = X_i(t) - \hat{X}_i(t)$  with a mean of $1.95 \times 10^{-6}$.}}
    \label{fig:15var}
\end{figure}

\section*{Appendix B}

This Appendix illustrates the constraints on the parameters implied by the unitary marginal variance of scores.

a) univariate model

\nin Let us suppose that the eigenscores $\beta_{i,j=1},..,\beta_{i,J_0}$ are independent and follow univariate AR($1$)-GARCH($1,1$) models. Then, the eigenscore $j$ on day $i$ is:

\begin{align*}
\beta_{i} &= \mu +  a \beta_{i-l} + \epsilon_{i}  \\
    \nu_{i} &= \varsigma_0 + \zeta \epsilon_{i-l}^2 +\varsigma \nu_{i-1}, 
\end{align*}

\nin where
$$\epsilon_{i} = \sqrt{v_{i}} z_{i}, \;\; \forall i=1,..,N \\ $$

\nin Then, the marginal variance $\sigma^2$ of $\epsilon_i$ satisfies:

 $$\sigma^2 = 1-a^2 = \varsigma_0/(1-(\zeta+\varsigma))$$

\nin where $\zeta+\varsigma \neq 1$ by the standard stationarity assumption on the GARCH(1,1).

\medskip

b) multivariate model

\begin{equation}
    \boldsymbol{\beta}_i = \mathbf{c}+\boldsymbol{\Pi}_1\boldsymbol{\beta}_{i-1} + \boldsymbol{\epsilon}_i
    \; i=1,...N
\end{equation}

\nin and the conditional variance of $\boldsymbol{\epsilon}_i$ is

\begin{equation}
    \mathbf{H}_i = \mathbf{CC}' + a \boldsymbol{\epsilon}_{i-1}\boldsymbol{\epsilon}'_{i-1} + g\mathbf{H}_{i-1}
    \; i=1,...N
\end{equation}

\nin Then the marginal variance $\Sigma$ of $\boldsymbol{\epsilon}_i$ satisfies

$$\Sigma = Id - \Pi \Pi' = CC'/(1-(a+g))$$

\nin where $a+g \neq 1$.

\medskip

\end{document}